%% file: bare_jrnl_new_sample4.tex
\documentclass[compsoc,journal]{IEEEtran}
\usepackage{amsmath,amsfonts}
\usepackage{algorithmic}
\usepackage{algorithm}
\usepackage{array}
\usepackage[caption=false,font=footnotesize]{subfig}
\usepackage{textcomp}
\usepackage{stfloats}
\usepackage{url}
\usepackage{verbatim}
\usepackage{graphicx}
\usepackage{cite}
\usepackage{booktabs}
\usepackage{multirow}
\usepackage{graphicx}
\usepackage{siunitx}
\usepackage{tabularx}
\begin{document}
\bstctlcite{IEEEtranBSTCTL_NoDash}

\title{SEROA: Stable Encrypted-Traffic Representation and Online Adaptation for Service-Aware Networks}

\author{Zongxi Liu,
  Shuai Mao,
  Jie Chuai,
  and~Hongyang Du$^*$

  \thanks{Zongxi Liu and Hongyang Du are with the Department of Electrical and Computer Engineering, The University of Hong Kong, Hong Kong SAR, China (e-mail: duhy@hku.hk).}
  \thanks{Shuai Mao and Jie Chuai are with Noah's Ark Lab, Huawei Technologies Ltd., Hong Kong SAR, China.}}
\markboth{}%
{Shell \MakeLowercase{\textit{et al.}}: A Sample Article Using IEEEtran.cls for IEEE Journals}


\maketitle

\begin{abstract}
  Mobile networks are evolving toward service-aware operation to support heterogeneous applications with diverse quality-of-service (QoS) and quality-of-experience (QoE) requirements. In practice, however, the widespread use of encryption makes service types difficult to observe directly, which limits the deployment of service-aware optimization in operational networks. Recent raw-byte representation learning methods have improved offline encrypted-traffic classification, but their robustness in realistic deployment remains limited due to temporal drift and the lack of explicit labels for online adaptation. This paper proposes Stable Encrypted-Traffic Representation and Online Adaptation (SEROA), a deployment-oriented framework for encrypted-traffic classification in service-aware mobile networks. In the offline stage, SEROA exploits protocol structure to learn temporally stable representations by applying masked modeling to stable fields, protocol-aware transformations and contrastive learning to unstable fields, and time-adversarial learning to suppress time-specific cues. In the online stage, SEROA uses feedback records containing QoE measurements, wireless states, and allocation decisions to update the deployed classifier through contextual-bandit learning. Experiments show that SEROA consistently outperforms representative pretrained baselines under both random and time-based evaluation protocols, and remains effective under QoE-driven online adaptation. These results demonstrate the value of stable representation learning and online adaptation for practical service-aware mobile networks.
\end{abstract}

\begin{IEEEkeywords}
  Encrypted traffic classification, online learning, temporal drift, representation learning.
\end{IEEEkeywords}

\input{sections/intro.tex}
\input{sections/related_works.tex}
\input{sections/Preliminaries.tex}
\input{sections/motivation.tex}
\input{sections/method.tex}
\input{sections/PG.tex}
\input{sections/experiments.tex}

\section{Discussion and Conclusion}
\subsection{Discussion}
This paper reports results on the public CSTNET datasets to ensure reproducibility and fair comparison. Beyond this public evaluation, the proposed method has also been tested on a private industrial dataset and deployed with an industrial collaborator. The dataset cannot be released because traffic traces are closely related to user privacy and operational confidentiality, which remains a fundamental constraint in network-trace sharing~\cite{Mohammady2021TraceAnonymization}.

Moreover, the current online-learning setting is still simplified. It considers service-aware power allocation with a minimum-rate-oriented QoE utility, while practical RAN optimization may involve additional resource dimensions such as subcarrier assignment and scheduling, together with richer QoE metrics beyond minimum rate~\cite{Sousa2020QoEOrientedScheduling,Elgabli2018QoERA,Mei2021RANSlicingDRL,Polese2023ColORAN}. Extending the framework to these settings is left for future work.

\subsection{Conclusion}
This paper presented a deployment-oriented framework for encrypted traffic classification under temporal drift. The offline component learns stable representations by separating protocol-structural and time-sensitive bytes, combining stable-field masked modeling, protocol-aware contrastive learning, and time-adversarial training. The online component uses feedback records containing QoE, wireless-state, and allocation information to update the classifier head without explicit service labels.

Experiments on CSTNET-APP and CSTNET-SERVICE show consistent gains over representative pretrained baselines under random and time-based splits, frozen and unfrozen fine-tuning, and QoE-driven online adaptation. These results show that stable representation learning and label-efficient online adaptation can improve encrypted traffic classification for service-aware mobile networks.


\bibliographystyle{IEEEtran}
\bibliography{refs}

\end{document}

%% file: sections/intro.tex
\section{Introduction}\label{sec:intro}
\IEEEPARstart{M}{obile} networks are evolving from the traditional
best-effort service model toward a service-aware paradigm, driven by
the rapid growth of latency-sensitive and experience-driven
applications such as artificial intelligence (AI) agents, eXtended Reality (XR), cloud gaming,
and live streaming~\cite{TS22261_ServiceReq,TR26928_XR5G,Laghari2019CloudGamingQoE,Bronzino2019EncryptedVideoQoE}.
To meet these heterogeneous requirements, 5G and
the emerging 5G-Advanced/6G architectures introduce fine-grained
quality-of-service (QoS) management mechanisms, including QoS Flows,
the 5G QoS Identifier (5QI), and the QoS Flow Identifier (QFI)
\cite{TS23501_QoS,TS23503_PCC}. The core network can configure QoS
parameters according to the service type, and the radio access network (RAN) can in
turn exploit this information for base station (BS) radio resource scheduling,
thereby improving the user-perceived quality-of-experience
(QoE)~\cite{Sousa2020QoEOrientedScheduling,Ojaghi2022SlicedRAN,Mei2021RANSlicingDRL,Liu2026LAMeTA}.

In practice, however, obtaining such service-type information has
become increasingly difficult because most operational traffic is now
encrypted. The widespread deployment of transport layer security (TLS)
has made encrypted traffic
increasingly ubiquitous in operational networks
\cite{rfc8446}. Encryption hides
application-layer payloads and most protocol semantics, rendering
traditional deep packet inspection (DPI) methods that rely on
plaintext content largely ineffective
\cite{Rezaei2019DeepLearningETC}.  As a
result, the service-type information assumed by many service-aware
optimization studies is not directly observable in encrypted mobile
networks, which limits the practical deployability of
service-aware resource management~\cite{Liu2026LAMeTA}. Consequently, accurately and robustly
identifying the service type of each flow under encryption has become
a central problem in service-aware mobile networking~\cite{Liu2026LAMeTA}.

To infer service types from encrypted traffic, recent
encrypted-traffic classification studies have shifted from manual
inspection and hand-crafted features to representation learning from
observable packet bytes.
Representative methods such as ET-BERT \cite{Lin2022ETBERT},
YaTC \cite{Zhao2023YaTC}, and
NetMamba \cite{Wang2024NetMamba} adapt
masked  modeling from natural language processing (NLP) and computer vision (CV) to
encrypted traffic, and achieve strong performance on offline
classification benchmarks. These results suggest that raw-byte traffic
contains exploitable signals for service identification, while
removing the dependence on hand-crafted features.
Despite this progress, existing raw-byte-based pretraining approaches
remain largely confined to idealized offline evaluation, and their
robustness in real mobile-network deployment remains limited
\cite{Navid2023DataDrift,Zhao2025SweetDanger,Wickramasinghe2025SoK}.
In deployment, a classifier is pretrained and fine-tuned on
historical traffic, and then applied to future flows whose
distribution shifts as applications and protocols evolve over time. This temporal drift can substantially degrade the accuracy
of encrypted-traffic classifiers~\cite{Taylor2018Robust,Navid2023DataDrift,Malekghaini2024AutoML4ETC}.
If the pretrained
encoder encodes time-specific shortcuts, fine-tuning has
limited ability to remove them, and temporal drift can induce larger
performance degradation~\cite{Zhao2025SweetDanger,Wickramasinghe2025SoK}.

We argue that this limitation mainly comes from two causes. The first
is field-agnostic pretraining. Existing methods directly transplant
NLP or CV objectives to encrypted traffic and treat all bytes as
homogeneous tokens or patches
\cite{Devlin2019BERT,He2022MAE,Lin2022ETBERT,Zhao2023YaTC}. Encrypted
traffic contains protocol-structural fields, environment-dependent
statistical fields, and session-specific or cryptographically opaque
fields, whose semantics and temporal stability differ substantially. In 
particular, IP, TCP, and TLS fields can have different stability properties. A uniform
pretraining objective can spend capacity reconstructing random
content or memorizing environment-specific patterns, yielding
representations that generalize poorly across time. The second cause
is insufficient support for online adaptation. When the deployment
distribution drifts substantially, historical offline training alone
cannot ensure accurate prediction for current traffic, and the
classifier must adapt to the new distribution.
Existing raw-byte pretraining studies largely stop at offline
evaluation and do not specify how a deployed model should learn
online in a mobile network. Such learning is difficult because
ground-truth service labels are usually unavailable after deployment.
The network can only log weak deployment signals, such as radio
conditions, scheduling outcomes, and QoE measurements, which are noisy
and indirectly related to class labels. Offline pretraining must
therefore produce representations robust enough to support lightweight
online updates under noisy weak feedback.

Motivated by these observations, we propose Stable Encrypted-Traffic
Representation and Online Adaptation (SEROA), a deployment-oriented
framework for encrypted-traffic classification in service-aware mobile
networks. SEROA first learns temporally stable traffic representations by
exploiting protocol structure and reducing the influence of
time-sensitive fields. It then supports lightweight online adaptation
after deployment by using feedback records containing QoE,
wireless-state, and allocation information. This design connects robust offline pretraining with online
service-aware resource allocation under temporal drift.

Our main contributions are summarized as follows.
\begin{itemize}
      \item We conduct a deployment-aware empirical analysis of
            representative raw-byte pretrained classifiers, showing that
            their strong random-split performance does not directly
            transfer to time-based testing or frozen-encoder adaptation.
            This performance gap motivates stable representations and
            online adaptation for practical deployment.
      \item We propose a protocol-aware offline pretraining framework that
            combines stable-field masked modeling,
            protocol-aware transformations, contrastive learning,
            and time-adversarial learning to learn representations that
            remain robust across time and effective under lightweight
            adaptation.
      \item We design a joint online feedback framework between the core network and the RAN,
            which uses classification-assisted resource allocation and
            feedback records containing QoE, wireless-state, and
            allocation information
            to update the deployed classifier through online
            contextual-bandit learning.
      \item We conduct extensive offline and online experiments on
            public CSTNET datasets. The results show consistent
            gains over representative pretrained baselines in both
            offline evaluation settings and under QoE-driven online
            adaptation.
\end{itemize}

The remainder of this paper is organized as follows.
Section~\ref{sec:related} reviews related work.
Section~\ref{sec:prelim_pipeline} formalizes the deployment
pipeline and the service-aware allocation model.
Section~\ref{sec:motivation} presents the deployment gap analysis.
Section~\ref{sec:method} introduces the proposed pretraining
method. Section~\ref{sec:pg} develops the online adaptation
mechanism, and Section~\ref{sec:experiments} reports the
experimental results.

%% file: sections/related_works.tex
\section{Related Work}\label{sec:related}

\subsection{Encrypted Traffic Classification}
Early network-traffic classifiers relied on port numbers and
DPI,
matching signatures against plaintext
headers or payloads. The deployment of TLS-based encryption has
reduced the visibility of application-layer
content, and shifted traffic classification toward side-channel
signals such as packet length, packet direction, timing, and early
flow behavior~\cite{moore2005toward, Crotti2007traffic,Bernaille2006early}.
Taylor et~al. showed that encrypted mobile applications can be
identified from statistical fingerprints extracted from traffic traces,
as demonstrated by AppScanner and a later robustness-oriented follow-up study~\cite{Taylor2018Robust,Taylor2016AppScanner}.
Lotfollahi et~al. and Rezaei et~al. then introduced end-to-end deep
models over raw packets or early-flow features, improving classification
accuracy under encrypted payloads~\cite{Lotfollahi2020DeepPacket,Rezaei2020MobileAppDL}. Other studies map
flows into intermediate representations, such as flowpic images,
and exploit convolutional, few-shot, or contrastive learning
architectures~\cite{Shapira2021FlowPic,Horowicz2022few}.
However, the
wide deployment of TLS and rapidly evolving application protocols
has reduced the reliability of such manually designed features. As
payload semantics are hidden and traffic patterns change across versions,
devices, and deployment environments, classifiers based on fixed
statistical descriptors become increasingly difficult to generalize.

To reduce the dependence on hand-crafted features, recent work has further moved from supervised deep models to raw-byte
self-supervised pretraining. Lin et~al. proposed ET-BERT, which learns contextual datagram
representations with masked language modeling~\cite{Lin2022ETBERT}.
Zhao et~al. introduced YaTC, which adapts masked autoencoding to traffic bytes~\cite{Zhao2023YaTC}.
Wang et~al. proposed NetMamba, which improves long-sequence traffic modeling with state-space
architectures~\cite{Wang2024NetMamba}. Chen et~al. and Liu et~al. developed
MIETT and FlowletFormer to model
packet-to-flow relations and protocol-aligned context for traffic
representation learning~\cite{Chen2025MIETT,liu2025FlowletFormer}.
These methods demonstrate the value of raw-byte representations for
encrypted-traffic classification. However, they treat packet bytes
as homogeneous tokens and adopt
pretraining objectives from NLP or computer vision. The heterogeneous stability properties of IP, TCP, and TLS
fields are therefore not explicitly modeled in these pretraining objectives.

\subsection{Drift Robustness}
Temporal drift is an important challenge for encrypted-traffic classification.
It can be induced by server-side deployment changes, device
heterogeneity, and evolving user behavior. Earlier mobile-app fingerprinting studies
reported accuracy variation across time, devices, and application
versions~\cite{Taylor2018Robust}. Recent evaluations further show that
random-split benchmarks can substantially overestimate deployment
performance when future traffic is chronologically separated from the
training data~\cite{Navid2023DataDrift,Zhao2025SweetDanger}. Malekghaini
et~al. studied classifier aging in mobile-network traffic and explored lightweight architectures for future data~\cite{Malekghaini2024AutoML4ETC}. Shapira et~al.'s FlowPic and related
systematization-of-knowledge (SoK)-style studies also report performance degradation under distribution
shift and emphasize the limitations of static legacy
benchmarks~\cite{Shapira2021FlowPic,Wickramasinghe2025SoK}.

These studies establish the importance of time-based evaluation and
drift-aware encrypted-traffic classification. However, most existing
analyses and mitigation methods are still built on flow-level
statistics, image-like flow descriptors, graph representations,
architecture search, or post-deployment model updates. Cui et~al.
proposed FG-SAT, which selects stable attributes from flow-graph
representations to reduce environmental sensitivity~\cite{Cui2025FGSAT}.
Chen et~al. proposed drift-oriented self-evolving encrypted-traffic
classification, which updates a deployed classifier with imperfect
labels~\cite{chen2025SelfEvolvingETC}, while Lin et~al. addressed
non-independent-and-identically-distributed (non-IID) traffic classification through self-supervised instruction
tuning in ETooL~\cite{Lin2025ETooL}. Despite improving robustness to distribution shifts, these methods are not suitable for raw-byte pretraining.

\subsection{Service-Aware Mobile Networking and Online Adaptation}
Service-aware mobile networking is supported by the 5G QoS
architecture. The 3GPP TS~23.501 defines QoS Flows, 5QI, and QFI as
standard mechanisms for mapping service requirements to forwarding
and access-side treatment~\cite{TS23501_QoS}. The next-generation
RAN (NG-RAN) then
binds QoS flows to radio resources according to the configured QoS
profiles~\cite{TS38300_NGRAN, TS23503_PCC}. In this context,
encrypted-traffic classification supplies the service type required
for QoS management and dynamic resource allocation. Fei et~al. studied
real-time encrypted-traffic classification in 5G non-standalone (NSA) networks as a
support module for QoS control and mobile-network resource
management~\cite{Fei2023RealTime, Rezaei2019DeepLearningETC}.

In parallel, QoE-oriented scheduling, application-aware resource
allocation, and slice-aware RAN optimization have developed into
active research directions. QoE-aware wireless scheduling studies
optimize radio resources for user-perceived
experience~\cite{Elgabli2018QoERA}.
Reinforcement learning and contextual-bandit methods have also been
used for RAN slicing, Open RAN (O-RAN) control, and cellular-network
configuration~\cite{Mei2021RANSlicingDRL,Polese2023ColORAN}.
Bronzino et~al. inferred streaming video quality from encrypted traffic,
and Dimopoulos et~al. measured video QoE from encrypted
traffic~\cite{Bronzino2019EncryptedVideoQoE,Dimopoulos2016EncryptedVideoQoE}.
These works show that service identification, QoE estimation, and online
wireless control are individually feasible. However, public studies rarely
couple them into a closed loop in which the classifier supports RAN
scheduling and RAN-side QoE feedback records are used to update the
deployed classifier itself.

Overall, prior work has advanced encrypted-traffic representation learning,
drift-aware evaluation, and service-aware mobile-network optimization
along largely separate directions. Existing raw-byte pretraining methods
mainly target offline classification, drift-robust methods are not
designed for raw-byte pretraining models, and service-aware networking studies
usually treat traffic classifiers as fixed inputs to resource allocation.
Therefore, the joint problem of temporally robust raw-byte representation
learning, QoE-driven RAN scheduling, and label-free online classifier
adaptation remains insufficiently studied. This paper addresses this gap
by incorporating protocol-field stability into pretraining and updating a
lightweight classifier head from RAN-side QoE feedback records after
deployment.

%% file: sections/Preliminaries.tex
\section{System Model and Problem Formulation}\label{sec:prelim_pipeline}
\begin{figure*}
  \centering
  \includegraphics[width=0.95\textwidth]{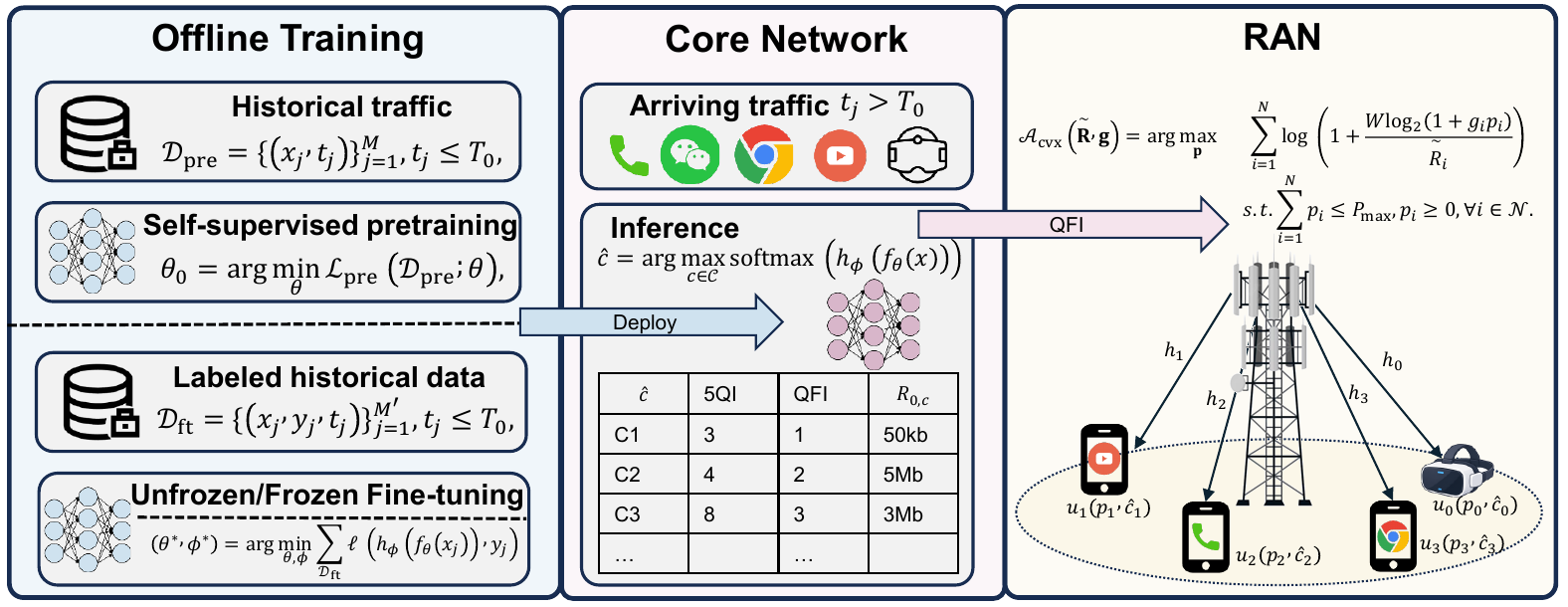}
  \caption{The pipeline of deploying an encrypted-traffic classification model in a mobile network.}
  \label{fig:scene}
\end{figure*}
\subsection{Deployment Pipeline for Encrypted-Traffic Representation}
Deploying an encrypted-traffic classification model in a mobile
network typically involves three stages: offline pretraining,
downstream fine-tuning, and online inference at deployment time as shown in Fig.~\ref{fig:scene}.

\paragraph{Pretraining} The network first
collects a large dataset of encrypted traffic with $M$ samples from
historical operations
\begin{equation}
  \mathcal{D}_{\mathrm{pre}}
  = \{(x_j,t_j)\}_{j=1}^{M},
  \quad t_j < T_0,
  \label{eq:prelim_dataset_pre}
\end{equation}
where $x_j$ is an encrypted-traffic sample, $t_j$ is its capture
time, and $T_0$ is the time at which the model is trained and ready
for deployment. Pretraining is carried out in a self-supervised
manner that does not rely on manual labels and learns
representations from past traffic:
\begin{equation}
  \theta_0
  = \arg\min_{\theta}
  \mathcal{L}_{\mathrm{pre}}
  \!\left( \mathcal{D}_{\mathrm{pre}};\theta \right),
  \label{eq:prelim_pretrain}
\end{equation}
where $f_{\theta}(\cdot)$ denotes the encoder and
$\mathcal{L}_{\mathrm{pre}}$ can take the form of masked modeling or other self-supervised objectives
\cite{Lin2022ETBERT,Zhao2023YaTC,Wang2024NetMamba}.

\paragraph{Fine-Tuning} After pretraining, the model is adapted to
the downstream classification task using a set of labeled
historical data with $M'$ samples
\begin{equation}
  \mathcal{D}_{\mathrm{ft}}
  = \{(x_j,y_j,t_j)\}_{j=1}^{M'},
  \quad t_j < T_0,
  \label{eq:prelim_dataset_ft}
\end{equation}
where $y_j \in \mathcal{C}$ denotes the service class or application type
of the flow
and $\mathcal{C}$ is the set of admissible services or applications.
Two typical fine-tuning strategies are considered. The first, \emph{unfrozen fine-tuning},
jointly updates the pretrained encoder and the classification
head:
\begin{equation}
  (\theta^\ast,\phi^\ast)
  = \arg\min_{\theta,\phi}
  \sum_{(x_j,y_j)\in \mathcal{D}_{\mathrm{ft}}}
  \ell\!\left( h_{\phi}(f_{\theta}(x_j)), y_j \right),
  \label{eq:prelim_unfrozen}
\end{equation}
where $h_{\phi}(\cdot)$ is the classification head and
$\ell(\cdot)$ is the cross-entropy loss. The second,
\emph{frozen fine-tuning}, keeps the encoder frozen and trains only
the lightweight classification head:
\begin{equation}
  \phi^\ast
  = \arg\min_{\phi}
  \sum_{(x_j,y_j)\in \mathcal{D}_{\mathrm{ft}}}
  \ell\!\left( h_{\phi}(f_{\theta_0}(x_j)), y_j \right),
  \quad \theta = \theta_0 .
  \label{eq:prelim_frozen}
\end{equation}
Although unfrozen full-parameter fine-tuning typically attains stronger
offline accuracy, it
degrades model stability and generalization. Therefore, both strategies
remain viable depending on the target objective and deployment scenario.

\paragraph{Deployment Inference} Once trained, the model is shipped
to inference modules co-located with the core network.
For each newly arriving encrypted-traffic sample $x \sim P_{t\ge T_0}(X)$, the
classifier forms the class posterior
\begin{equation}
  \boldsymbol{\xi}^{\mathrm{cls}}(x)
  = \mathrm{softmax}\!\left( h_{\phi}\!\left( f_{\theta}(x) \right) \right),
  \label{eq:prelim_posterior}
\end{equation}
and emits the predicted service class
$\hat{c} = \arg\max_{c\in\mathcal{C}} \xi^{\mathrm{cls}}_c(x)$.
$x$ denotes an early-flow sample constructed
from the initial packets of a newly arriving flow.
The classifier is invoked once, and the resulting service prediction is retained throughout the flow lifetime for QoS mapping and scheduling.

\subsection{Service-Aware Resource Allocation}
\label{sec:prelim_qoe}
Consider a single-cell downlink system in which the BS
serves a set of active users $\mathcal{N} = \{1,2,\ldots,N\}$.
Each user $i \in \mathcal{N}$ is associated with one application
flow whose true service class is $c_i \in \mathcal{C}$. The
downlink transmission from the BS to user $i$ is modeled as
\begin{equation}
  z_i = \sqrt{p_i}\,h_i a_i + n_i,\quad h_i = \sqrt{\beta_i}\,\tilde{h}_i
  \label{eq:prelim_tx}
\end{equation}
where $z_i$ is the received baseband signal at user $i$, $p_i$ is the transmit power allocated to user $i$, $a_i$ is the unit-power transmitted symbol, $n_i \sim \mathcal{CN}(0,\sigma^2)$ is additive white Gaussian noise, and $\mathcal{CN}$ denotes the circularly symmetric complex Gaussian distribution.
The channel $h_i$ follows Rayleigh fading~\cite{goldsmith2005wireless}, $\tilde{h}_i \sim \mathcal{CN}(0,1)$ is small-scale fading and $\beta_i$ is large-scale path loss.
When the BS allocates transmit power $p_i$ to user $i$, the
achievable downlink rate is
\begin{equation}
  \quad g_i = \frac{|h_i|^2}{\sigma^2}, r_i = W\log_2(1+g_i p_i).
  \label{eq:prelim_rate}
\end{equation}
where $g_i$ denotes the normalized channel gain, $r_i$ is the
achievable downlink rate of user $i$, and $W$ is the system bandwidth.
The predicted service class $\hat{c}$ is used by the core-network QoS
control logic to select the treatment of the corresponding service data
flow. Under policy control function (PCF) policy, the session management
function (SMF) provides the QoS Flow context with the assigned QFI and
5QI to the RAN over N2, while the user plane function (UPF) marks N3 packets with the QFI,
enabling the RAN to map the flow to radio resources according to the
received QoS context~\cite{TS23501_QoS,TS23503_PCC,TS38300_NGRAN}.
We abstract the class-dependent effect to a reference
rate $R_{0,c}$, defined as the baseline throughput required for a
minimally acceptable experience of service class $c$.
Motivated by the Weber--Fechner
law~\cite{Reichl2010WeberFechnerQoE}, we adopt a logarithmic QoE model for each user $i$
with service class $c_i$:
\begin{equation}
  \begin{aligned}
    u_i(p_i,c_i)
     & = \log\!\left( 1 + \frac{r_i}{R_{0,c_i}} \right)                 \\
     & = \log\!\left( 1 + \frac{W\log_2(1+g_i p_i)}{R_{0,c_i}} \right).
  \end{aligned}
  \label{eq:prelim_qoe}
\end{equation}
The inner ratio $r_i/R_{0,c_i}$ normalizes the radio-side rate by
the service requirement, while the outer logarithm captures
diminishing marginal returns.

The RAN performs power allocation using the service information
available on the QoS path. Let
$\tilde{\mathbf{R}}=[\tilde{R}_1,\ldots,\tilde{R}_N]$ denote the
reference rates used by the RAN for the active flows. For a fixed channel vector
$\mathbf{g}=[g_1,\ldots,g_N]$ and a fixed reference-rate vector
$\tilde{\mathbf{R}}$, define the convex allocation operator
\begin{equation}
  \begin{aligned}
    \mathcal{A}_{\mathrm{cvx}}(\tilde{\mathbf{R}},\mathbf{g})
    = \arg\max_{\mathbf{p}}\quad
     & \sum_{i=1}^{N}
    \log\!\left(1+\frac{W\log_2(1+g_i p_i)}{\tilde{R}_i}\right) \\
    \mathrm{s.t.}\quad
     & \sum_{i=1}^{N} p_i \le P_{\max},\quad
    p_i \ge 0,\ \forall i \in \mathcal{N}.
  \end{aligned}
  \label{eq:prelim_alloc_operator}
\end{equation}
where $P_{\max}$ is a total transmit-power budget. For fixed $\tilde{\mathbf{R}}$, each term in the objective is concave
in $p_i$, and the feasible set is convex. Hence
\eqref{eq:prelim_alloc_operator} can be solved directly by standard
convex optimization methods.

In current classification-assisted deployment, the core-network
classifier emits the predicted classes
$\hat{\mathbf{c}}=[\hat{c}_1,\ldots,\hat{c}_N]$. The RAN then uses Predicted-Class allocation:
\begin{equation}
  \tilde{R}^{\mathrm{pred}}_i = R_{0,\hat{c}_i},\qquad
  \mathbf{p}^{\mathrm{pred}}
  = \mathcal{A}_{\mathrm{cvx}}(\tilde{\mathbf{R}}^{\mathrm{pred}},\mathbf{g}).
  \label{eq:prelim_pred_alloc}
\end{equation}
The realized user experience is still governed by the true service
classes through $\sum_i u_i(p^{\mathrm{pred}}_i,c_i)$. Therefore,
classification accuracy directly affects the mismatch between the
allocation objective used by the RAN and the QoE perceived by users.
When $\hat{c}_i=c_i$ for all active flows, the predicted-class
allocation coincides with the Oracle allocation :
\begin{equation}
  \tilde{R}^{\mathrm{orc}}_i = R_{0,c_i},\qquad
  \mathbf{p}^{\mathrm{orc}}
  = \mathcal{A}_{\mathrm{cvx}}(\tilde{\mathbf{R}}^{\mathrm{orc}},\mathbf{g}),
  \label{eq:prelim_oracle_alloc}
\end{equation}
which serves as an ideal upper reference for
classification-assisted allocation.

Additionally, in some deployment and evaluation settings, the allocation module may
not have access to reliable per-flow service-class information. For
example, in LTE evolved packet core (EPC)
deployments, the RAN-side scheduler may not receive an end-to-end QFI/5QI for each
encrypted flow~\cite{TS23501_QoS,TS23503_PCC,TS38300_NGRAN,TS37340_MRDC}.
In this low-information setting, all flows are assigned the same
class-mix reference rate with a Class-Agnostic allocation:
\begin{equation}
  \bar{R}_0
  = \frac{1}{|\mathcal{C}|}\sum_{c\in\mathcal{C}} R_{0,c},
  \qquad
  \mathbf{p}^{\mathrm{agn}}
  = \mathcal{A}_{\mathrm{cvx}}(\bar{R}_0\mathbf{1}_N,\mathbf{g}),
  \label{eq:prelim_agn_alloc}
\end{equation}
where $\mathbf{1}_N$ denotes the $N$-dimensional all-one vector.

%% file: sections/motivation.tex
\section{Deployment Gap Analysis}\label{sec:motivation}

Building on the deployment setting defined in
Section~\ref{sec:prelim_pipeline}, this section examines whether
existing pretrained encrypted-traffic models remain effective when
applied to future traffic. We evaluate three representative methods, including
ET-BERT~\cite{Lin2022ETBERT}, YaTC~\cite{Zhao2023YaTC}, and
NetMamba~\cite{Wang2024NetMamba}, on the public CSTNET
dataset~\cite{Lin2022ETBERT}, where all three were originally reported
to achieve Macro-F1 scores above $90\%$.

\subsection{Dataset Construction and Evaluation Protocols}
\label{sec:mot_setup}

\begin{figure}[t]
    \centering
    \subfloat[Per-label sample time distribution for the CSTNET]{%
        \includegraphics[width=\columnwidth]{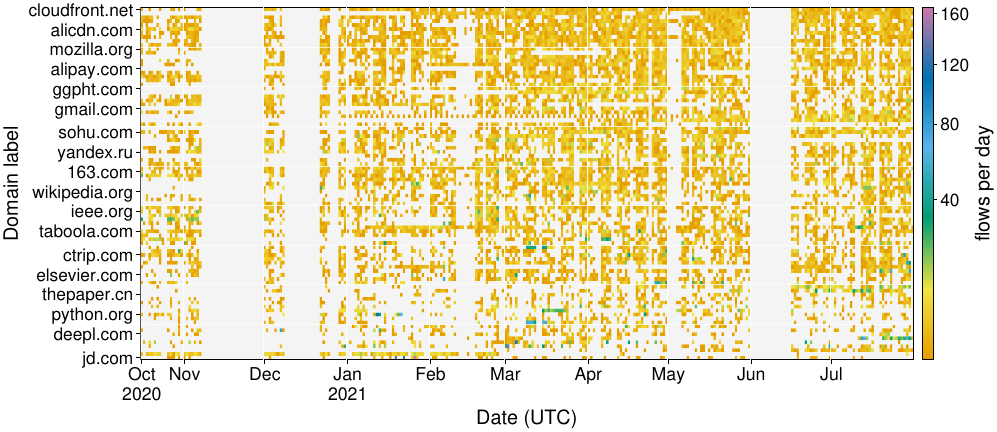}%
        \label{fig:label_day_heatmap}}\\
    \subfloat[Two evaluation protocols on CSTNET-SERVICE.]{%
        \includegraphics[width=\columnwidth]{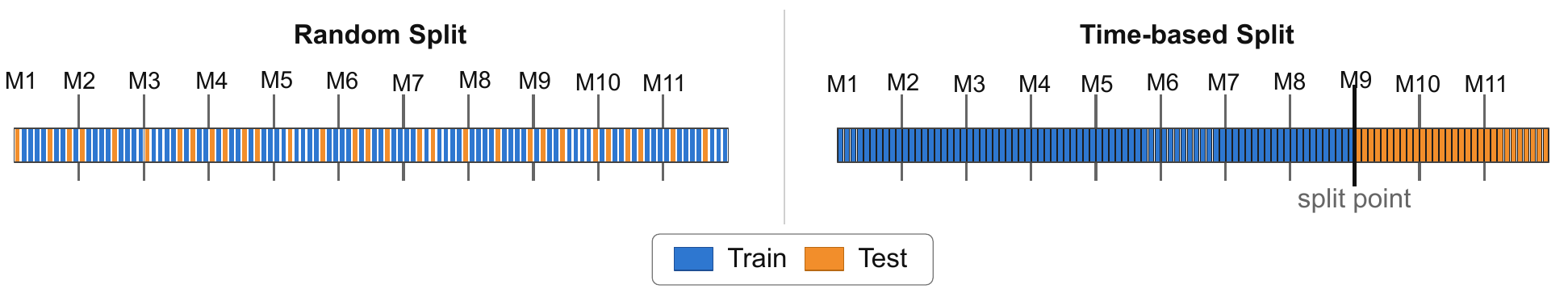}%
        \label{fig:split}}
    \caption{Temporal characteristics of the CSTNET dataset and
        the two evaluation protocols compared in this work.}
    \label{fig:dataset_split}
\end{figure}

CSTNET is a public TLS~1.3 encrypted-traffic dataset whose flows
are annotated with fine-grained application-type labels~\cite{Lin2022ETBERT}.
The collection window spans roughly eleven months, and the per-label
arrival pattern is shown in Fig.~\ref{fig:label_day_heatmap}. To reduce
confounding effects from limited support and uneven temporal coverage,
we retain application types with sufficient samples and active days,
adequate presence on both sides of the chronological boundary, and no
excessive concentration on a single day. This procedure retains $89$
of the original $120$ application types and forms CSTNET-APP. We then
aggregate the retained application types by service semantics into five
categories $\mathcal{C}=\{C_1,\ldots,C_5\}$ to form CSTNET-SERVICE.
Table~\ref{tab:cstnet_service_profiles} summarizes their service
profiles, standardized 5QI values, and reference rates $R_{0,c}$ used
in the online evaluation~\cite{TS23501_QoS,ITU_G1010,
    AppleHLSAuthoring,RFC6817_LEDBAT}.

\begin{table}[!t]
    \centering
    \footnotesize
    \caption{Service profiles used to construct CSTNET-SERVICE and perform online evaluation.}
    \label{tab:cstnet_service_profiles}
    \begin{tabularx}{\linewidth}{@{}c
            >{\raggedright\arraybackslash}X
            cc@{}}
        \toprule
        \textbf{Class} & \textbf{Service category}                    & \textbf{5QI} &
        \textbf{$R_{0,c}$}                                                                         \\
        \midrule
        $C_1$          & Interactive low-rate services                & 3            & $50$\,kbps  \\
        $C_2$          & Lightweight information-access services      & 6            & $24$\,kbps  \\
        $C_3$          & High-rate media and content services         & 4            & $5$\,Mbps   \\
        $C_4$          & General data services                        & 8            & $3$\,Mbps   \\
        $C_5$          & Background support and distribution services & 9            & $100$\,kbps \\
        \bottomrule
    \end{tabularx}
\end{table}

We evaluate CSTNET-SERVICE under two protocols. The first is the conventional
\emph{random split}, adopted by prior work~\cite{Lin2022ETBERT,
    Zhao2023YaTC,Wang2024NetMamba}, shuffles all flows and partitions them
into training and test sets with a $7{:}3$ ratio. The other is the deployment-oriented
\emph{time-based split} uses the chronological threshold
$T_0=\texttt{2021-05-06}$. Samples collected before $T_0$ form the
training set, and those collected on or after $T_0$ form the test set.
Detailed cleaning criteria, the domain-to-service mapping,
and exact split indices are provided in the implementation.
\footnote{The implementation are provided in the supplementary material.} The two protocols are illustrated
in Fig.~\ref{fig:split}.

\subsection{Existing Methods Degrade under Time-Based Split}
\label{sec:mot_degrade}

Fig.~\ref{fig:accuracy_drop} reports the Macro-F1 of ET-BERT,
YaTC, and NetMamba on CSTNET-SERVICE under random
split with unfrozen fine-tuning, time-based split with unfrozen
fine-tuning, and time-based split with frozen fine-tuning. Under the conventional random-split protocol, all three
methods exceed $90\%$ Macro-F1, in agreement with their original
reports. However, when the protocol is changed to the time-based split,
the Macro-F1 of all three methods drops to roughly $75\%$. When
the backbone is further frozen,
all three methods collapse to $20\%\text{--}50\%$ Macro-F1.
This indicates that the high accuracy reported by these raw-bytes pretrained models
relies simultaneously on a temporally independent and identically distributed (i.i.d.) test set, which is
violated in the real deployment scenario.
\subsection{Confirming Temporal Drift via Feature Divergence}
\label{sec:mot_drift}

We further measure how strongly the train and test sets differ in
protocol-level features. For each flow, we extract a set $\mathcal{F}$
of protocol-aware features from its packet trace, including flow-level
statistics and commonly used fields from IP and TCP headers, TLS
records, and TLS handshakes. The complete feature definitions and
extraction procedure are provided in the accompanying implementation.

\begin{figure}[t]
    \centering
    \subfloat[Macro-F1 of prior works on
        CSTNET-SERVICE under three different settings.]{%
        \includegraphics[width=0.48\columnwidth]{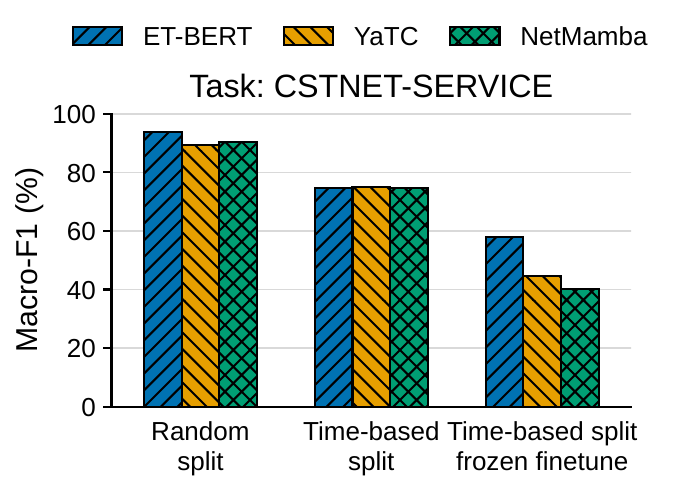}%
        \label{fig:accuracy_drop}}%
    \hfill
    \subfloat[Per-class mean Jensen--Shannon divergence
        $D^{s}_{c}$ under time-based split versus random split.]{%
        \includegraphics[width=0.48\columnwidth]{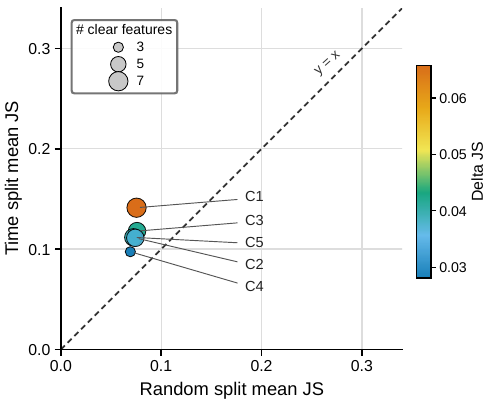}%
        \label{fig:js_drift}}
    \caption{Empirical evidence that chronological evaluation is
        substantially harder than random-split evaluation.}
    \label{fig:time_drift_gap}
\end{figure}

For each service class $c \in \mathcal{C}$, each feature
$f \in \mathcal{F}$, and each split protocol $s \in
    \{\mathrm{time}, \mathrm{random}\}$, we estimate the empirical
train and test distributions $P^{s}_{c,f}$ and $Q^{s}_{c,f}$ and
quantify their disagreement via the Jensen--Shannon (JS)
divergence~\cite{Lin1991JSD}:
\begin{equation}
    JS^{s}_{c,f}
    = \tfrac{1}{2}\mathrm{KL}\!\left(P^{s}_{c,f}\,\big\|\,M^{s}_{c,f}\right)
    + \tfrac{1}{2}\mathrm{KL}\!\left(Q^{s}_{c,f}\,\big\|\,M^{s}_{c,f}\right),
    \label{eq:mot_js}
\end{equation}
where $M^{s}_{c,f} = (P^{s}_{c,f} + Q^{s}_{c,f})/2$. Averaging
over features gives a service-level drift score
$D^{s}_{c} = |\mathcal{F}|^{-1}\sum_{f\in\mathcal{F}} JS^{s}_{c,f}$,
and the gap induced by chronological evaluation is
\begin{equation}
    \Delta_c \;=\; D^{\mathrm{time}}_{c} - D^{\mathrm{random}}_{c}.
    \label{eq:mot_delta}
\end{equation}
Fig.~\ref{fig:js_drift} plots $D^{\mathrm{time}}_c$ against
$D^{\mathrm{random}}_c$ for the five service categories. All five lie
strictly above the diagonal $y=x$, i.e. $\Delta_c>0$
for every class. These observations confirm that temporal drift
exists in CSTNET-SERVICE and changes the protocol-level
features used by the pretrained models.

\subsection{Source of the Degradation}
\label{sec:mot_insight}

The key limitation of existing raw-byte pretraining methods is their
field-agnostic design. These methods directly inherit
token- or patch-level objectives from NLP and CV
\cite{Devlin2019BERT,He2022MAE,Lin2022ETBERT,Zhao2023YaTC,Wang2024NetMamba}
and apply them uniformly to encrypted traffic bytes. As a result, all
sampled bytes are optimized as if they had the same protocol role and
temporal stability.

In practice, encrypted-traffic fields exhibit heterogeneous statistical
properties. For example, a TLS \texttt{session\_id} is generated for a
specific connection and is therefore essentially random. \texttt{ip.ttl}
is coupled to routing and collection environments. And packet length is
comparatively stable because it reflects protocol and application-level
structure. Motivated by this observation, we partition the feature
vocabulary $\mathcal{F}$ into three field sets: protocol-structural
stable fields $\mathcal{F}_{\mathrm{stab}}$, environment-dependent
fields $\mathcal{F}_{\mathrm{env}}$, and session-specific ephemeral
fields $\mathcal{F}_{\mathrm{eph}}$. Table~\ref{tab:protocol_field_taxonomy}
summarizes the corresponding assignment within this feature vocabulary.

Therefore, an effective encrypted-traffic pretraining method should
explicitly account for field heterogeneity. It should exploit stable
protocol structure while reducing reliance on environment- and
session-specific variation.

\begin{table}[!t]
    \centering
    \small
    \caption{Field sets used for protocol-aware traffic modeling.}
    \label{tab:protocol_field_taxonomy}
    \begin{tabularx}{\linewidth}{@{}p{0.13\linewidth}
            >{\raggedright\arraybackslash}X@{}}
        \toprule
        \textbf{Field set} & \textbf{Included fields} \\
        \midrule

        $\mathcal{F}_{\mathrm{stab}}$
                           &
        Packet length, TLS record type, TLS handshake type,
        and other protocol-structural fields.         \\

        \addlinespace

        $\mathcal{F}_{\mathrm{env}}$
                           &
        \texttt{ip.\{tos, dsfield, ttl\}},
        \texttt{tcp.window\_size}.                    \\

        \addlinespace

        $\mathcal{F}_{\mathrm{eph}}$
                           &
        \texttt{ip.id}, \texttt{tcp.\{seq, ack\}}, \texttt{tls.handshake.\{random, session\_id, key\_share\}},
        \texttt{tls.app\_data}.                       \\

        \bottomrule
    \end{tabularx}
\end{table}

%% file: sections/method.tex
\section{Stable Offline Representation Learning}\label{sec:method}
\begin{figure*}
    \centering
    \includegraphics[width=0.99\textwidth]{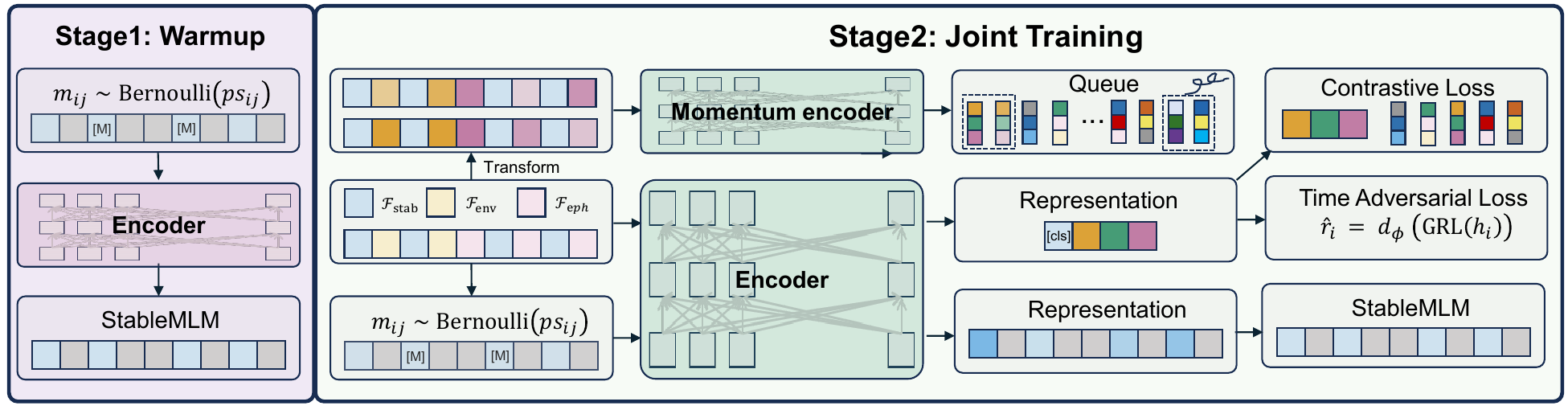}
    \caption{Overview of the proposed protocol-aware pretraining method.}
    \label{fig:method}
\end{figure*}

Motivated by the insight in Section~\ref{sec:mot_insight}, we propose a
\emph{protocol-aware} pretraining method that handles stable,
and unstable information separately. Our method uses the same encoder as
ET-BERT~\cite{Lin2022ETBERT} for fair comparison, but with a
different pretraining method. It comprises three modules:
(i) stable masked language modeling (StableMLM) (Section~\ref{sec:method_smlm}), which performs masked
reconstruction only on stable fields. (ii) Protocol-aware
Contrastive Learning (Section~\ref{sec:method_cl}), which
improves the stability of the learned representations. And
(iii) Time-Adversarial Module
(Section~\ref{sec:method_time}), which suppresses residual
time-correlated information at the representation level.

\subsection{Stable Masked Byte Modeling}\label{sec:method_smlm}
We use the field sets
$\mathcal{F}_{\mathrm{stab}}$, $\mathcal{F}_{\mathrm{env}}$, and
$\mathcal{F}_{\mathrm{eph}}$ defined in
Section~\ref{sec:mot_insight}. To connect field-level categories to
byte-level inputs, let $\mathcal{I}_i(\mathcal{S})\subseteq\{1,\ldots,L\}$
denote the byte positions occupied in $x_i$ by fields in a field set
$\mathcal{S}$. We attach to each sample $x_i$ of length $L$ a
\emph{stability mask}
$\mathbf{b}_i = [b_{i1},\ldots,b_{iL}] \in \{0,1\}^L$ defined as
\begin{equation}
    b_{ij} \;=\; \mathbb{1}\!\left[\,j \in
        \mathcal{I}_i(\mathcal{F}_{\mathrm{stab}})\,\right],
    \label{eq:method_stability_mask}
\end{equation}
which means only positions belonging to protocol-structural stable
fields are flagged $b_{ij}=1$.

StableMLM performs masked byte reconstruction
only on stable positions, so the encoder learns
byte-level embeddings from stable protocol structure instead of
spending capacity on random content. Given $\mathbf{b}_i$, we sample the MLM
mask with probability $p_{\mathrm{mask}}\in[0,1]$ only over stable positions:
\begin{equation}
    m_{ij} \sim \mathrm{Bernoulli}(p_{\mathrm{mask}} b_{ij}).
    \label{eq:method_mask_constraint}
\end{equation}
Masked positions are replaced with a \texttt{[MASK]} token to form
the corrupted input $\tilde{x}_i$, and the encoder output is fed
to a reconstruction head $\zeta_\psi(\cdot)$ that predicts the
original byte $\hat{\pi}^{\mathrm{byte}}_{ij} = \zeta_\psi\!\left( f_\theta(\tilde{x}_i)_j \right).$
Here $\hat{\pi}^{\mathrm{byte}}_{ij}$ denotes the categorical distribution predicted over the byte vocabulary at position $j$ of sample $i$, and $\hat{\pi}^{\mathrm{byte}}_{ij}(x_{ij})$ is the probability assigned to the original byte $x_{ij}$.
The StableMLM objective is the cross-entropy averaged over masked
positions in a mini-batch of size $B$:
\begin{equation}
    \mathcal{L}_{\mathrm{SMLM}}
    = -\,\frac{1}{\sum_{i=1}^{B}\sum_{j=1}^{L} m_{ij}}
    \sum_{i=1}^{B} \sum_{j=1}^{L}
    m_{ij}\,\log\hat{\pi}^{\mathrm{byte}}_{ij}(x_{ij}).
    \label{eq:method_smlm_loss}
\end{equation}

\subsection{Protocol-aware Contrastive Learning}
\label{sec:method_cl}

StableMLM teaches the encoder to predict stable bytes from their
context, but does not guarantee that the resulting representation
is invariant to changes on unstable fields. We make this
invariance explicit through a contrastive objective built on a
protocol-aware augmentation.

\paragraph{Protocol-aware Transform}
The positive view $x_i^{+}$ keeps every stable byte of $x_i$
($b_{ij}=1$) intact and redraws unstable fields.
Let $v_{i,f}$ and $v_{i,f}^{+}$ denote the values of field $f$ in
the original sample $x_i$ and the augmented view $x_i^{+}$, respectively.
For each session-specific ephemeral field $f \in \mathcal{F}_{\mathrm{eph}}$,

\begin{equation}
    v_{i,f}^{+} \;\sim\; \mathrm{Uniform}(\mathcal{V}_f),
    \label{eq:method_eph_random}
\end{equation}
where $\mathcal{V}_f$ is the admissible value space of $f$.
For ordinary ephemeral fields, sampling is independent across
fields. For stateful ephemeral fields such as $\texttt{tcp.seq}$ and $\texttt{tcp.ack}$,
we instead apply one shared
per-flow transformation to all coupled fields to preserve
intra-flow consistency.
For each environmental field $f \in \mathcal{F}_{\mathrm{env}}$,

\begin{equation}
    v_{i,f}^{+} \;\sim\; \hat{P}(v_f),
    \label{eq:method_resample}
\end{equation}
which means $v_{i,f}^{+}$ is sampled from the
distribution of field $f$ estimated from the pretraining set.
Because fields in $\mathcal{F}_{\mathrm{env}}$ encode both environment-specific statistics
and temporal noise, sampling from $\hat{P}(v_f)$
preserves the former while randomizing the latter.
Additionally, checksums that depend on the modified bytes are recomputed and length-related
parsing boundaries are kept intact, so
$x_i^{+}$ is a syntactically valid sample that differs from $x_i$
precisely on
$\mathcal{I}_i(\mathcal{F}_{\mathrm{eph}} \cup \mathcal{F}_{\mathrm{env}})$.

\paragraph{Contrastive Objective}
We feed the transformed pairs into a contrastive objective to
obtain a better representation, and instantiate it with the MoCo
v2~\cite{He2020MoCo,Chen2020MoCov2} framework. Under this framework,
the encoder $f_\theta$ trained in StableMLM continues to be
updated by back-propagation and plays the role of the
\emph{query encoder}. Additionally, we introduce a
\emph{momentum encoder} $f_{\bar\theta}$ that shares the
architecture of $f_\theta$ and whose parameters track $\theta$
through an exponential moving average (EMA),
\begin{equation}
    \bar\theta \;\leftarrow\; \mu\,\bar\theta + (1-\mu)\,\theta,
    \quad \mu \in [0,1).
    \label{eq:method_ema}
\end{equation}
With $\mu$ close to $1$, $\bar\theta$ evolves slowly, so the keys
produced by $f_{\bar\theta}$ over many past steps remain mutually
consistent and can be reused as negatives.
The encoder $f_\theta$
embeds the original view and $f_{\bar\theta}$ embeds the augmented
view. Each branch is followed by a multilayer perceptron (MLP) projection head
denoted by $\zeta_q$ and $\zeta_k$, respectively, and
$\ell_2$ normalization:
\begin{equation}
    q_i = \mathrm{norm}\!\left(
    \zeta_q\!\left( \mathrm{Pool}( f_\theta(x_i) ) \right)
    \right),
    \label{eq:method_query}
\end{equation}
\begin{equation}
    k_i^{+} = \mathrm{norm}\!\left(
    \zeta_k\!\left( \mathrm{Pool}( f_{\bar\theta}(x_i^{+}) ) \right)
    \right),
    \label{eq:method_key}
\end{equation}
where $\mathrm{Pool}(\cdot)$ is a sequence-pooling operator. To decouple the number
of negatives from the batch size, we maintain a first-in
first-out queue $\mathcal{Q}$ of recently produced keys with
capacity $|\mathcal{Q}|$. At each
step, the current-batch keys $\{k_i^{+}\}_{i=1}^{B}$ are enqueued
and the oldest entries are evicted, so $\mathcal{Q}$ caches a
large negative pool accumulated over many past steps without
enlarging the batch.

Given a mini-batch of queries $\{q_i\}_{i=1}^{B}$, positive keys
$\{k_i^{+}\}_{i=1}^{B}$, and the negative keys $\{k_l\}$ in
$\mathcal{Q}$, the information noise-contrastive estimation (InfoNCE) loss~\cite{Oord2018CPC} is
\begin{equation}
    \mathcal{L}_{\mathrm{CL}}
    = -\,\frac{1}{B} \sum_{i=1}^{B}
    \log \frac{\exp(q_i^{\!\top} k_i^{+}/\tau)}
    {\exp(q_i^{\!\top} k_i^{+}/\tau)
        + \sum_{k_l\in\mathcal{Q}} \exp(q_i^{\!\top} k_l/\tau)},
    \label{eq:method_infonce}
\end{equation}
where $\tau>0$ is a temperature hyperparameter. Because $x_i^{+}$ differs from $x_i$ only on positions in
$\mathcal{I}_i(\mathcal{F}_{\mathrm{eph}} \cup \mathcal{F}_{\mathrm{env}})$,
minimizing $\mathcal{L}_{\mathrm{CL}}$ is equivalent to enforcing
that the encoder be insensitive to perturbations of ephemeral and
environment-dependent bytes, while remaining free to exploit the
information carried by $\mathcal{F}_{\mathrm{stab}}$.

\subsection{Time-Adversarial Module}
\label{sec:method_time}

The two modules above rely on manually specified protocol priors.
As a complement, we add an end-to-end implicit branch that lets
the encoder suppress residual time-correlated information.

We linearly normalize the capture time $t_i$ of each pretraining
sample to $\tilde{t}_i \in [0,1]$. Let
$\mathbf{z}_i = \mathrm{Pool}( f_\theta(x_i) )$ denote the pooled sequence
representation. We route $\mathbf{z}_i$ through a gradient reversal layer
(GRL)~\cite{Ganin2016DANN} into a time-regression head $d_\nu(\cdot)$:
\begin{equation}
    \hat{t}_i \;=\; d_\nu\!\left( \mathrm{GRL}(\mathbf{z}_i) \right),
    \label{eq:method_time_head}
\end{equation}
and supervise the prediction with a Smooth-$\ell_1$ loss:
\begin{equation}
    \mathcal{L}_{\mathrm{time}}
    = \frac{1}{B} \sum_{i=1}^{B}
    \mathrm{SmoothL1}\!\left( \hat{t}_i,\, \tilde{t}_i \right).
    \label{eq:method_time_loss}
\end{equation}

The GRL acts as the identity in the forward pass but reverses the
gradient propagated to the encoder during back-propagation. The
resulting adversarial objective is
\begin{equation}
    \min_{\nu}\max_{\theta}\;
    \mathcal{L}_{\mathrm{time}}(\theta,\nu),
    \label{eq:method_time_game}
\end{equation}
where $d_\nu$ learns to recover the capture time, whereas $f_\theta$
learns to remove time-predictive information from $\mathbf{z}_i$.

\subsection{Training Pipeline}\label{sec:method_pipeline}

Both the contrastive objective and the day-regression branch
operate on the pooled representation $\mathrm{Pool}(f_\theta(\cdot))$.
If $f_\theta$ has not yet acquired informative byte-level features,
the gradients of both objectives are computed on a representation
without discriminative structure and carry little useful signal.
We therefore adopt a \emph{two-stage} pretraining schedule.

\paragraph{StableMLM Warm-up}
We first optimize $\mathcal{L}_{\mathrm{SMLM}}$ alone for $T_w$ epochs, so that
$f_\theta$ acquires byte-level embeddings and contextual
representations on stable positions, providing an effective
initialization for the subsequent two objectives.

\paragraph{Joint Training}
Starting from the warmed-up encoder, all three objectives are
optimized jointly for $T_j$ epochs:
\begin{equation}
    \mathcal{L}_{\mathrm{total}}
    \;=\; \mathcal{L}_{\mathrm{SMLM}}
    \;+\; \lambda_c\,\mathcal{L}_{\mathrm{CL}}
    \;+\; \lambda_t\,\mathcal{L}_{\mathrm{time}},
    \label{eq:method_total_loss}
\end{equation}
where $\lambda_c,\lambda_t \ge 0$ are loss weights.
The complete training procedure is summarized in Algorithm~\ref{alg:method}.

\paragraph{Stage~3: Fine-Tuning}
The pretrained encoder $f_\theta$ is then used directly in the two
downstream configurations defined in
Section~\ref{sec:prelim_pipeline}, namely the unfrozen and frozen
fine-tuning given by \eqref{eq:prelim_unfrozen} and
\eqref{eq:prelim_frozen}.

\begin{algorithm}[t]
    \caption{Protocol-aware Pretraining for Time-Drift Encrypted Traffic Classification}
    \label{alg:method}
    \begin{algorithmic}[1]
        \REQUIRE pretraining set $\mathcal{D}_{\mathrm{pre}}=\{(x_i,t_i)\}$; encoder $f_\theta$; MLP heads $\zeta_\psi$, $\zeta_q,\zeta_k$; time head $d_\nu$.
        \STATE \textit{// Stage~1: StableMLM warm-up}
        \FOR{$\iota=1,\ldots,T_w$}
        \STATE sample $\{x_i\}_{i=1}^{B}\sim\mathcal{D}_{\mathrm{pre}}$; build $\mathbf{b}_i,m_i,\tilde{x}_i$ via \eqref{eq:method_stability_mask}--\eqref{eq:method_mask_constraint}
        \STATE update $(\theta,\psi)$ along $\nabla\mathcal{L}_{\mathrm{SMLM}}$ in \eqref{eq:method_smlm_loss}
        \ENDFOR
        \STATE $\bar\theta\gets\theta$;\ $\zeta_k\gets \zeta_q$;\ fill $\mathcal{Q}$ with keys produced by $f_{\bar\theta}$
        \STATE \textit{// Stage~2: joint pretraining}
        \FOR{$\iota=1,\ldots,T_j$}
        \STATE sample $\{(x_i,t_i)\}_{i=1}^{B}\sim\mathcal{D}_{\mathrm{pre}}$
        \STATE build $\mathbf{b}_i,m_i,\tilde{x}_i$ via \eqref{eq:method_stability_mask}--\eqref{eq:method_mask_constraint}
        \STATE build positive view $x_i^{+}$ via \eqref{eq:method_eph_random}--\eqref{eq:method_resample}
        \STATE compute $\mathcal{L}_{\mathrm{SMLM}}$ on $\tilde{x}_i$ via \eqref{eq:method_smlm_loss}
        \STATE compute $q_i$ and $k_i^{+}$ via \eqref{eq:method_query}--\eqref{eq:method_key}
        \STATE compute $\mathcal{L}_{\mathrm{CL}}$ with negatives in $\mathcal{Q}$ via \eqref{eq:method_infonce}
        \STATE normalize $t_i$ to $\tilde{t}_i\!\in\![0,1]$;
        \STATE compute $\hat{t}_i$ via \eqref{eq:method_time_head}
        \STATE compute $\mathcal{L}_{\mathrm{time}}$ via \eqref{eq:method_time_loss}
        \STATE $\mathcal{L}_{\mathrm{total}}\gets\mathcal{L}_{\mathrm{SMLM}}+\lambda_c\mathcal{L}_{\mathrm{CL}}+\lambda_t\mathcal{L}_{\mathrm{time}}$
        \STATE update $(\theta,\psi,\zeta_q,\nu)$ along $\nabla\mathcal{L}_{\mathrm{total}}$\ \textit{// GRL flips the sign of $\nabla_\theta\mathcal{L}_{\mathrm{time}}$}
        \STATE $\bar\theta\!\gets\!\mu\bar\theta\!+\!(1\!-\!\mu)\theta$;\ $\zeta_k\!\gets\!\mu\zeta_k\!+\!(1\!-\!\mu)\zeta_q$
        \STATE enqueue $\{k_i^{+}\}$ into $\mathcal{Q}$; evict oldest entries
        \ENDFOR
        \STATE \textit{// Stage~3: fine-tuning}
        \STATE \textbf{return} $f_\theta$; downstream fine-tune via \eqref{eq:prelim_unfrozen} or \eqref{eq:prelim_frozen}
    \end{algorithmic}
\end{algorithm}

%% file: sections/PG.tex
\section{Closed-Loop Online Adaptation}\label{sec:pg}
Section~\ref{sec:method} provides an offline training method that is
more robust to temporal drift. A more accurate classifier enables the
RAN to select reference rates closer to the true service requirements,
thereby improving the QoE of the power allocation in
Section~\ref{sec:prelim_qoe}. However, when the deployment
distribution drifts severely, the predicted service classes may still
deviate from the latent services. The resulting allocation then
deviates from the Oracle allocation in
\eqref{eq:prelim_oracle_alloc}. This gap motivates a closed-loop
online learning framework that uses QoE feedback to adapt the deployed
classifier and improve the allocation performance simultaneously.
\subsection{Closed-Loop Formulation}\label{sec:pg_model}

Online learning is difficult in deployment because explicit service
labels are unavailable. However, QoE feedback $u_i(p_i,c_i)$ can serve as a weak label.
Although it is
affected by the wireless channel and the applied allocation, it still
reflects whether the selected service treatment matches the latent
service. We therefore use QoE-based weak feedback to update the
deployed classifier in a closed loop.

The closed loop requires two additional information paths. The first
path carries the classifier posterior from the core-network inference
function to the RAN-side allocation module. In the standard QoS
procedure, QFI and 5QI provide the RAN with a hard QoS treatment,
whereas online adaptation requires posterior confidence over candidate
services. We therefore expose $\boldsymbol{\xi}^{\mathrm{cls}}$ as a
flow-level soft service descriptor. This descriptor can be carried as
operator-internal AI/ML
control metadata, without changing the user-plane QoS marking or the
QFI and 5QI mapping procedure. Although current specifications do not
define this posterior as a standard QoS field, its use is consistent
with 3GPP AI/ML data-collection studies~\cite{TR37817_RANAI}.

The second path provides the feedback required for classifier-head
adaptation. The allocation decision is made at the RAN, whereas
$h_\phi$ is maintained in the core network. Updating $h_\phi$ therefore
requires the gradient signal induced by RAN-side allocation outcomes.
Since current 3GPP QoS and RAN-control interfaces do not specify
transport of neural-network gradients, the RAN reports an
update-equivalent record from which the core-network learner can
reconstruct the objective. The update-equivalent record contains
deployment feedback such as QoE measurements and RAN-side execution
information. Such records can be collected through multi-access edge
computing (MEC) or operations, administration, and maintenance (OAM)
data-collection paths~\cite{TS28405_QMC,ETSI_MEC012_RNIS,TR37817_RANAI}.
If model parameters or policy updates are exchanged, they can be handled
through 3GPP management-plane AI/ML model management. This design is
consistent with 3GPP AI/ML management support for model training,
testing, loading, update, and inference reporting~\cite{TS28105_AIML}.

With these two paths, the deployed system can be modeled as an
online expected-QoE maximization problem. At each decision epoch,
the core-network classifier produces the posterior
$\boldsymbol{\xi}^{\mathrm{cls}}_{i,\phi}$ for each active flow, and
the RAN observes the wireless state $g_i$. The allocation rule
$\mathcal{A}_\omega$ then maps these quantities to a feasible power
vector. The joint optimization problem is
\begin{equation}
    \begin{aligned}
        \max_{\phi,\,\omega}\quad
         & J(\phi,\omega)=\mathbb{E}\!\left[ U^{\mathrm{QoE}} \right]=
        \mathbb{E}\!\left[ \sum_{i=1}^{N} u_i(p_i, c_i) \right]        \\
        \mathrm{s.t.}\quad
         & \mathbf{p}
        = \mathcal{A}_\omega\!\left(
        \{(g_i,\boldsymbol{\xi}_{i,\phi}^{\mathrm{cls}})\}_{i=1}^{N}
        \right),                                                       \\
         & \sum_{i=1}^{N} p_i \le P_{\max},\quad p_i \ge 0,\quad
        \forall i\in\mathcal{N}.
    \end{aligned}
    \label{eq:pg_joint}
\end{equation}
This problem is nonconvex because the classifier posterior is nonlinear,
and the allocation rule
$\mathcal{A}_\omega$ is optimized jointly with this posterior
input.

\subsection{Policy-Gradient Solution}\label{sec:pg_solve}
\begin{figure}[!t]
    \centering
    \includegraphics[width=0.95\linewidth]{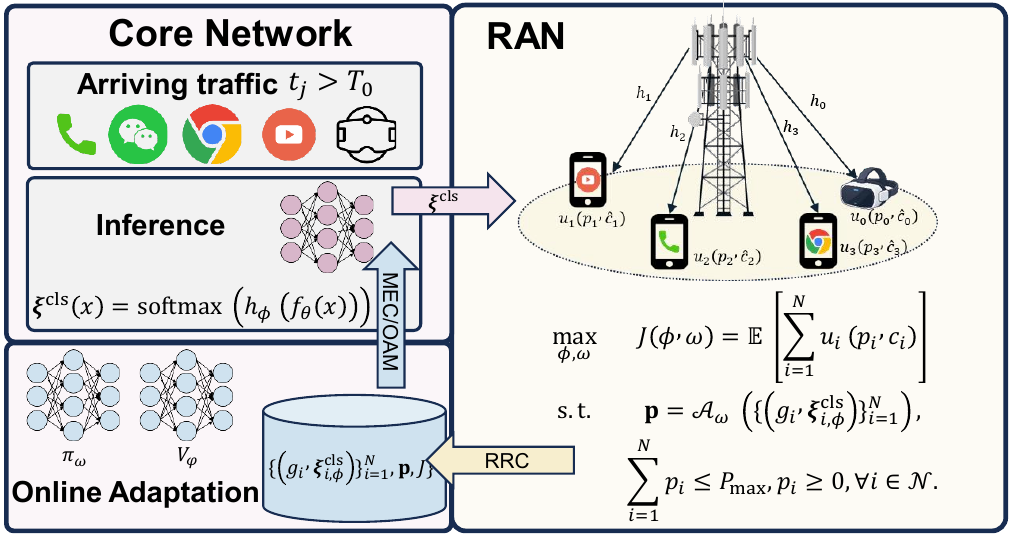}
    \caption{The online learning framework.}
    \label{fig:pg}
\end{figure}
To solve \eqref{eq:pg_joint}, we instantiate $\mathcal{A}_\omega$ with
a stochastic policy $\pi_\omega$ and optimize its parameters jointly
with the classifier head. Since
each flow arrival is a single decision without a state transition, we
cast the problem as a
contextual bandit~\cite{Li2010ContextualBandit} and solve it by policy
gradient. Let $\gamma_i=\log(1+g_i)$ and normalize it within each
rollout as
$\bar{\gamma}_i=(\gamma_i-\mu_\gamma)/(\sigma_\gamma+\varepsilon_{\mathrm{norm}})$,
where $\mu_\gamma$ and $\sigma_\gamma$ are the rollout mean and
standard deviation and $\varepsilon_{\mathrm{norm}}>0$ is a small
constant. The context state concatenates the normalized log channel
gain with the classifier posterior,
\begin{equation}
    \mathbf{s}^{\mathrm{ctx}}_{i,\phi}
    = \left[\,\bar{\gamma}_i,\; \boldsymbol{\xi}^{\mathrm{cls}}_{i,\phi}\,\right],\label{eq:pg_state}
\end{equation}
and the system context is
$\mathbf{s}^{\mathrm{ctx}}_{\phi}=[\mathbf{s}^{\mathrm{ctx}}_{1,\phi},\ldots,\mathbf{s}^{\mathrm{ctx}}_{N,\phi}]$.
The policy outputs
Dirichlet concentrations $\boldsymbol{\alpha}_\omega(\mathbf{s}^{\mathrm{ctx}}_{\phi})$ and
samples the power allocation
\begin{equation}
    \boldsymbol{\rho}\sim \mathrm{Dir}(\boldsymbol{\alpha}_\omega(\mathbf{s}^{\mathrm{ctx}}_{\phi})),\qquad p_i = P_{\max}\,\rho_i,\label{eq:pg_action}
\end{equation}
which meets the power budget by construction. The RAN executes the
sampled allocation and returns a rollout record to the core-network
learner. The rollout buffer $\mathcal{M}$ stores the context, the
sampled power vector, the behavior log probability, the policy
version or parameter snapshot, and the measured aggregate QoE reward
$U^{\mathrm{QoE}}$.

Let $\pi_\omega(\mathbf{p}\mid\mathbf{s}^{\mathrm{ctx}}_{\phi})$ denote the Dirichlet policy. We maximize the expected reward in \eqref{eq:pg_joint} by REINFORCE~\cite{Williams1992REINFORCE}, whose policy-gradient component is
\begin{equation}
    \nabla_\omega J =
    \mathbb{E}\!\left[\,\nabla_\omega \log\pi_\omega(\mathbf{p}\mid\mathbf{s}^{\mathrm{ctx}}_{\phi})
        U^{\mathrm{QoE}}\,\right].\label{eq:pg_grad}
\end{equation}
The aggregate QoE reward varies strongly with service type and channel realization, so we subtract a value-network baseline $V_\varphi(\mathbf{s}^{\mathrm{ctx}}_{\phi})$ and use the advantage
\begin{equation}
    A = U^{\mathrm{QoE}} - V_\varphi(\mathbf{s}^{\mathrm{ctx}}_{\phi}),\label{eq:pg_adv}
\end{equation}
normalized to zero mean and unit variance within each rollout, where $V_\varphi$ is trained with $\mathcal{L}_{\mathrm{val}}=\mathbb{E}[(V_\varphi(\mathbf{s}^{\mathrm{ctx}}_{\phi})-U^{\mathrm{QoE}})^2]$. To reuse each rollout under the limited online sample budget, we adopt the proximal policy optimization (PPO) clipped surrogate~\cite{Schulman2017PPO}. Let $\eta(\omega,\phi)$ denote the importance ratio between the current posterior-dependent policy and the behavior policy stored during rollout collection:
\begin{equation}
    \begin{aligned}
        \eta(\omega,\phi)
         & =
        \frac{\pi_\omega(\mathbf{p}\mid \mathbf{s}^{\mathrm{ctx}}_{\phi})}
        {\pi_{\omega_{\mathrm{old}}}(\mathbf{p}\mid
            \mathbf{s}^{\mathrm{ctx}}_{\phi_{\mathrm{old}}})}.
    \end{aligned}
    \label{eq:pg_ratio}
\end{equation}
Here $\omega_{\mathrm{old}}$ and $\phi_{\mathrm{old}}$ denote the
behavior policy and classifier-head parameters used to collect the
rollout.
\begin{equation}
    \mathcal{L}_{\mathrm{pol}} = -\mathbb{E}\!\left[\min\!\left(\eta(\omega,\phi)\,A,\;\mathrm{clip}(\eta(\omega,\phi),\,1-\epsilon,\,1+\epsilon)\,A\right)\right],\label{eq:pg_pol}
\end{equation}
where $\epsilon>0$ is the PPO clipping parameter that limits each update to a small trust region. The collected QoE rewards are fixed during the $K$ update epochs. With an entropy regularizer $\mathcal{H}(\pi_\omega)$ for exploration, the total loss is
\begin{equation}
    \mathcal{L} = \mathcal{L}_{\mathrm{pol}} + \lambda_v\,\mathcal{L}_{\mathrm{val}} - \lambda_e\,\mathcal{H}(\pi_\omega).\label{eq:pg_loss}
\end{equation}
The coefficients $\lambda_v\ge 0$ and $\lambda_e\ge 0$
are the weight of the value-function regression
term and the entropy regularization term, respectively.
Since the context state \eqref{eq:pg_state} contains
the posterior $\boldsymbol{\xi}^{\mathrm{cls}}_{i,\phi}$,
the core-network learner back-propagates $\mathcal{L}$ into the local
classifier head when replaying the feedback record and the corresponding
policy information.
Because the observed QoE reward is corrupted by channel
randomness and user-side estimation noise, updating the full encoder
could distort the stable representation learned offline. We therefore
freeze $f_{\theta_0}$ and update only $\{\omega,\varphi,\phi\}$.
After each update round, the refreshed policy parameters are
synchronized to the RAN-side allocation module through the same
management path.
Algorithm~\ref{alg:pg} summarizes the procedure.

\begin{algorithm}[t]
    \caption{Closed-Loop Bandit Adaptation}
    \label{alg:pg}
    \begin{algorithmic}[1]
        \REQUIRE frozen encoder $f_{\theta_0}$; classifier head $h_\phi$; policy $\pi_\omega$; value network $V_\varphi$.
        \STATE $\omega_{\mathrm{old}} \gets \omega$;\ $\phi_{\mathrm{old}} \gets \phi$
        \FOR{each rollout}
        \STATE initialize rollout buffer $\mathcal{M}$
        \FOR{each online decision epoch in the rollout}
        \STATE obtain classifier posteriors $\boldsymbol{\xi}^{\mathrm{cls}}_{i,\phi_{\mathrm{old}}}$ for active flows
        \STATE combine the posteriors with channel gains to form $\mathbf{s}^{\mathrm{ctx}}_{\phi_{\mathrm{old}}}$ via \eqref{eq:pg_state}
        \STATE allocate power $\mathbf{p}$ using $\pi_{\omega_{\mathrm{old}}}$ via \eqref{eq:pg_action} and record the behavior log probability and policy version
        \STATE apply $\mathbf{p}$ and store the QoE reward, channel gains, applied power, behavior log probability, and policy version or snapshot in $\mathcal{M}$
        \ENDFOR
        \STATE compute and normalize advantages in $\mathcal{M}$ via \eqref{eq:pg_adv}
        \FOR{$k=1,\ldots,K$}
        \STATE recompute current posteriors and states for transitions in $\mathcal{M}$
        \STATE evaluate $\mathcal{L}$ via \eqref{eq:pg_ratio}--\eqref{eq:pg_loss}
        \STATE update $\{\omega,\varphi,\phi\}$ along $\nabla\mathcal{L}$; keep $f_{\theta_0}$ frozen
        \ENDFOR
        \STATE synchronize the updated policy parameters $\omega$ to the RAN-side allocation module
        \STATE $\omega_{\mathrm{old}} \gets \omega$;\ $\phi_{\mathrm{old}} \gets \phi$
        \ENDFOR
    \end{algorithmic}
\end{algorithm}

For implementation, the RAN reports the
measured QoE reward, the wireless state , the applied power vector, the behavior log
probability through MEC/OAM data-collection
paths~\cite{TS28405_QMC,ETSI_MEC012_RNIS,TR37817_RANAI}.
And the policy network parameter updates can be exchanged through management-plane AI/ML
model management~\cite{TS28105_AIML}.

%% file: sections/experiments.tex
\section{Experiments}\label{sec:experiments}

\subsection{Experimental Setup}
\label{sec:exp_setup}

\textbf{Hardware and Implementation.}
All experiments were conducted on the same Ubuntu 24.04
server using PyTorch 2.8.0 and CUDA 12.8, with a 38-core x86-64
CPU and data-center GPUs, each providing 80 GB of memory and 19.5
TFLOPS peak FP32 performance.

\textbf{Datasets and protocols.}
We evaluate on CSTNET-APP and CSTNET-SERVICE under the random and
time-based splits defined in Section~\ref{sec:mot_setup}. For both
splits, we report unfrozen and frozen fine-tuning and use classification
accuracy (ACC) and Macro-F1 as the main metrics.

\subsection{Offline Evaluation}
\label{sec:exp_offline}

The offline pretraining hyperparameters of the proposed method are
summarized in Table~\ref{tab:ours_offline_hparams}. Our model uses the
same encoder architecture as ET-BERT and changes only the pretraining
objective. We use AdamW with batch size 32, learning rate
$5\times10^{-5}$, and weight decay 0.02. The learning rate is warmed
up from $5\times10^{-6}$ for 10 epochs and then cosine-decayed to
$1\times10^{-5}$ over a 50-epoch optimizer schedule. StableMLM is
optimized alone for 30 epochs, followed by 50 epochs of joint training.

We compare against ET-BERT~\cite{Lin2022ETBERT},
YaTC~\cite{Zhao2023YaTC}, and NetMamba~\cite{Wang2024NetMamba} using
their released weights and official fine-tuning protocols. ET-BERT is
fine-tuned with AdamW for 50 epochs using batch size 32 and learning
rate $6\times10^{-5}$ and its classifier learning rate in the frozen
setting is $2\times10^{-4}$. YaTC is fine-tuned with AdamW for 200
epochs using batch size 128 and base learning rate $2\times10^{-3}$.
NetMamba is fine-tuned with AdamW for 120 epochs using batch size 128
and base learning rate $2\times10^{-3}$.

\begin{table}[t]
    \centering
    \caption{Hyperparameters of the proposed offline method.}
    \label{tab:ours_offline_hparams}
    \small
    \setlength{\tabcolsep}{6pt}
    \renewcommand{\arraystretch}{1.08}
    \begin{tabular}{lc}
        \toprule
        \textbf{Hyperparameter}              & \textbf{Value} \\
        \midrule
        Mask probability $p_{\mathrm{mask}}$ & 0.3            \\
        EMA coefficient $\mu$                & 0.995          \\
        InfoNCE temperature $\tau$           & 0.07           \\
        MoCo queue capacity $|\mathcal{Q}|$  & 8192           \\
        Contrastive loss weight $\lambda_c$  & 10.0           \\
        Time-adversarial weight $\lambda_t$  & 5.0            \\
        StableMLM warm-up epochs $T_w$       & 30             \\
        Joint training epochs $T_j$          & 50             \\
        \bottomrule
    \end{tabular}
\end{table}

\textbf{Time-based split.}
Table~\ref{tab:time_results} reports the performance comparison
under the time-based split. Our method attains the best ACC and
Macro-F1 on both datasets and in both fine-tuning settings. On
CSTNET-APP, Macro-F1 improves from 38.48\%/16.33\%/7.41\% to
67.89\% in the frozen setting and from 69.88\%/68.65\%/68.28\%
to 72.14\% in the unfrozen setting. On CSTNET-SERVICE, it
improves from 50.03\%/27.71\%/20.75\% to 73.02\% in the frozen
setting and from 74.56\%/74.90\%/74.54\% to 82.19\% in the
unfrozen setting. The larger gain in the frozen setting indicates
stronger transfer under temporal drift.

\textbf{Random split.}
Table~\ref{tab:random_results} reports the performance comparison
under the random split. Our method also leads in this setting. On CSTNET-APP,
frozen Macro-F1 increases from 48.31\%/24.93\%/13.93\% to
86.85\%, and on CSTNET-SERVICE it increases from
55.12\%/27.66\%/23.86\% to 90.47\%. The unfrozen results are best
in every column. The result confirms a consistent gain on the
i.i.d.\ benchmark, which means our method is also effective in
conventional benchmarks without temporal drift.

\textbf{Ablation study.}
Table~\ref{tab:ablation_time} reports the ablation results under
the time-based setting. StableMLM alone lifts frozen Macro-F1 to 57.34\% on CSTNET-APP
and 66.48\% on CSTNET-SERVICE. Adding the contrastive branch
raises it to 64.41\% and 72.14\%, and adding the time-adversarial
branch raises it to 60.05\% and 69.77\%. The full model reaches
67.89\% and 73.02\%. These results show complementary
contributions from the three objectives.

\begin{figure*}[!t]
    \centering
    \subfloat[ET-BERT]{\includegraphics[width=0.245\textwidth]{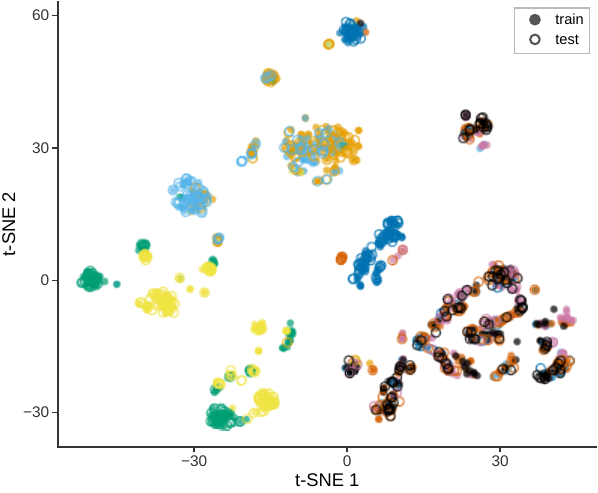}}\hfill
    \subfloat[YaTC]{\includegraphics[width=0.245\textwidth]{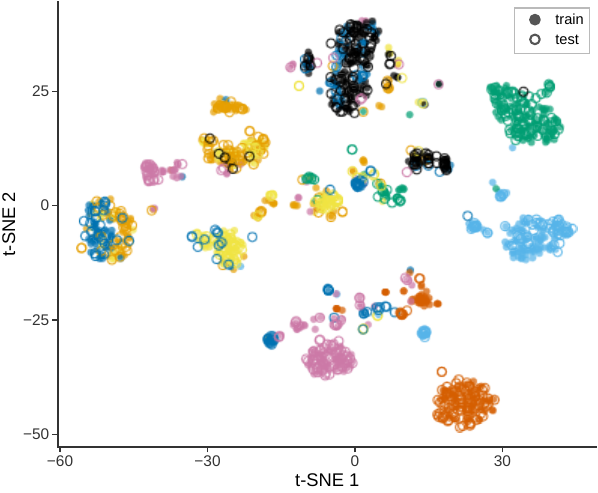}}\hfill
    \subfloat[NetMamba]{\includegraphics[width=0.245\textwidth]{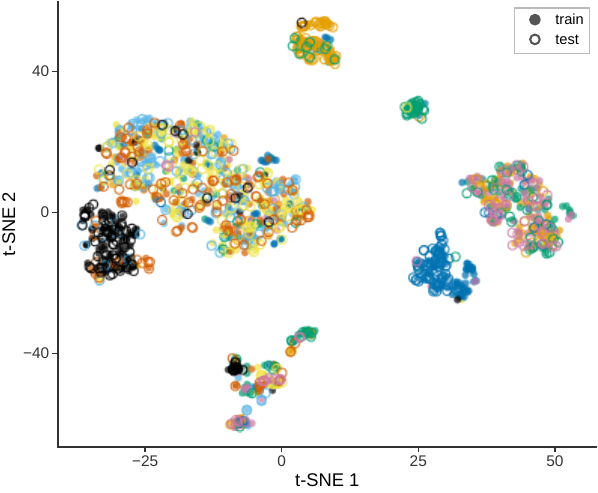}}\hfill
    \subfloat[Ours]{\includegraphics[width=0.245\textwidth]{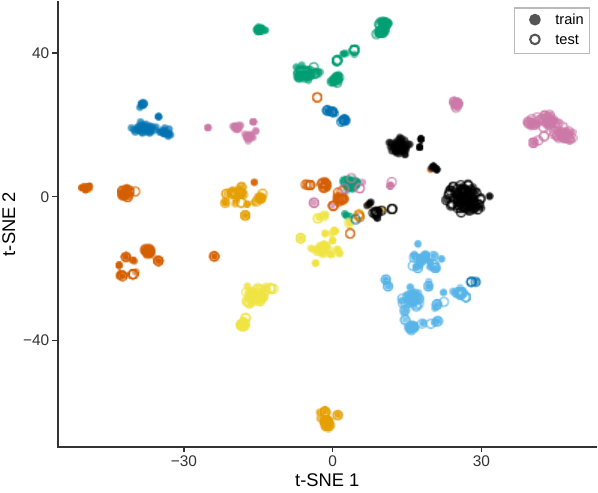}}\\[2pt]
    \subfloat[Ours w/o Time Adv]{\includegraphics[width=0.245\textwidth]{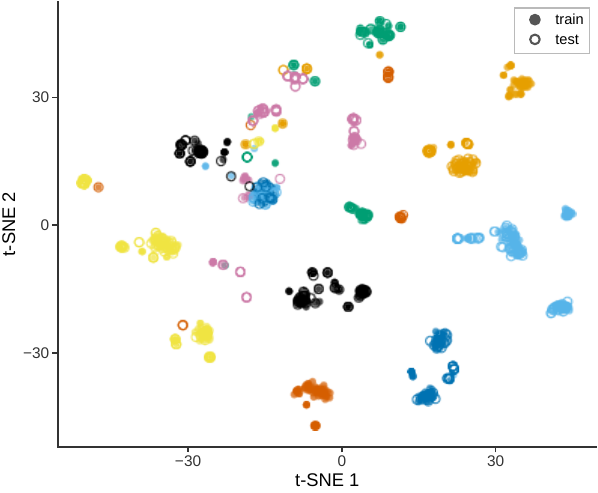}}\hfill
    \subfloat[Ours w/o Contrastive]{\includegraphics[width=0.245\textwidth]{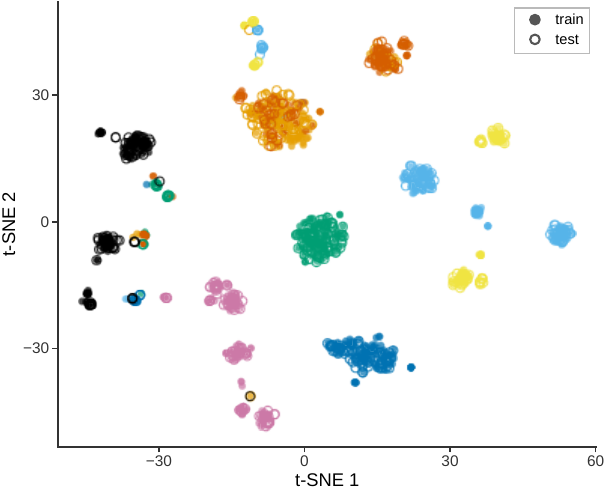}}\hfill
    \subfloat[Ours w/o Both]{\includegraphics[width=0.245\textwidth]{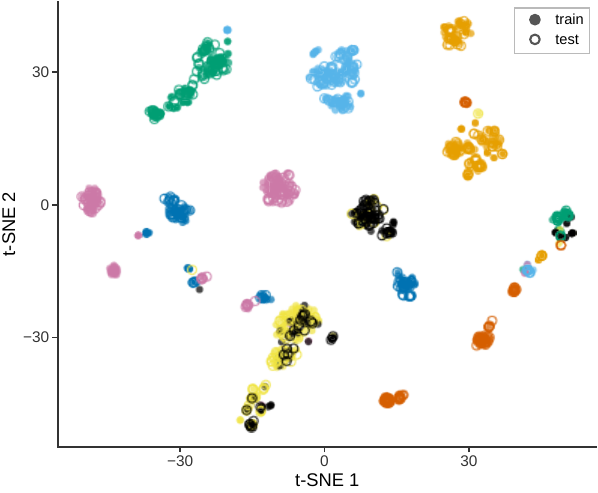}}\hfill
    \subfloat[5-NN Purity]{\includegraphics[width=0.245\textwidth]{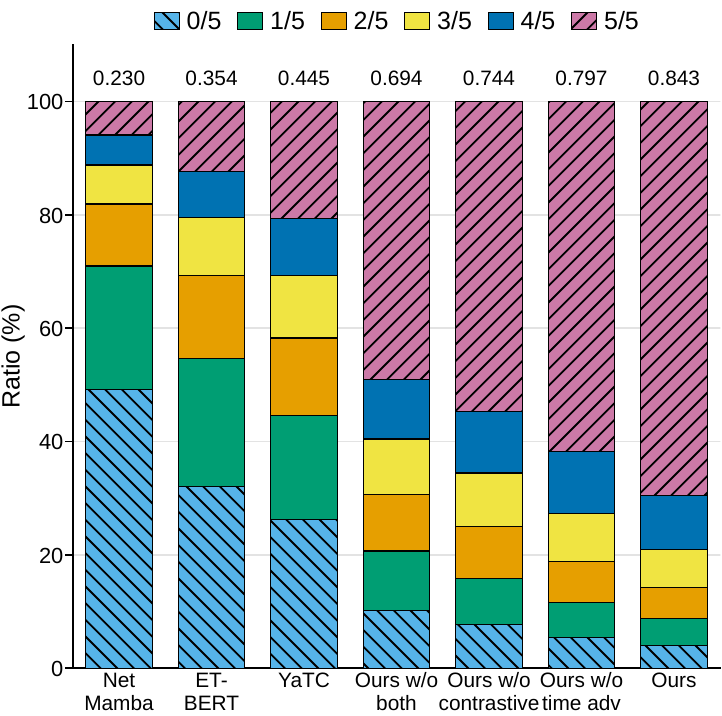}}
    \caption{Representation visualization on the cross-time test split. (a)--(c) are baselines, (d) is the proposed StableMLM+Contrastive+Time-Adversarial method, (e)--(g) ablate one or both components of (d), and (h) reports the 5-NN class purity for all methods.}
    \label{fig:rep_visual}
\end{figure*}

\textbf{Representation visualization.}
Figure~\ref{fig:rep_visual} visualizes the pooled encoder outputs
with t-distributed stochastic neighbor embedding (t-SNE) for seven randomly selected service classes from the
cross-time test split.
ET-BERT, YaTC, and NetMamba show overlapping clusters, whereas the
full model yields compact class-wise clusters. The ablated models
occupy intermediate positions. The five-nearest-neighbor (5-NN) purity further provides a
quantitative view of this local structure. For each sample, it
counts how many of its five nearest neighbors in the representation
space share the same class label. Larger values therefore indicate
stronger local class consistency. The baseline distributions are
dominated by the 0/5 bin, meaning that most samples have no
same-class neighbor among their five nearest neighbors. This
suggests that the released pretraining provides limited class
separation. By contrast, the full model
is concentrated in the 5/5 bin, showing that the proposed
pretraining learns locally coherent and class-discriminative
representations. 
The ablation variants perform between the baselines and the full model. Removing either the contrastive or time-adversarial objective reduces high-purity mass, while StableMLM alone also underperforms the full model.
These trends confirm complementary
contributions from all three pretraining components.

\begin{table*}[!t]
    \centering
    \caption{Performance comparison under the time-based setting.}
    \label{tab:time_results}
    \small
    \setlength{\tabcolsep}{5.5pt}
    \renewcommand{\arraystretch}{1.15}
    \begin{tabular*}{\textwidth}{@{\extracolsep{\fill}}l
            S[table-format=2.2] S[table-format=2.2]
            S[table-format=2.2] S[table-format=2.2]
            S[table-format=2.2] S[table-format=2.2]
            S[table-format=2.2] S[table-format=2.2]
            @{}}
        \toprule
        \multirow{3}{*}{Method}
        & \multicolumn{4}{c}{CSTNET-APP(time-based split)}
        & \multicolumn{4}{c}{CSTNET-SERVICE(time-based split)} \\
        \cmidrule(lr){2-5} \cmidrule(lr){6-9}
        & \multicolumn{2}{c}{Unfrozen}
        & \multicolumn{2}{c}{Frozen}
        & \multicolumn{2}{c}{Unfrozen}
        & \multicolumn{2}{c}{Frozen} \\
        \cmidrule(lr){2-3} \cmidrule(lr){4-5}
        \cmidrule(lr){6-7} \cmidrule(lr){8-9}
        & {ACC (\%)} & {Macro-F1 (\%)}
        & {ACC (\%)} & {Macro-F1 (\%)}
        & {ACC (\%)} & {Macro-F1 (\%)}
        & {ACC (\%)} & {Macro-F1 (\%)} \\
        \midrule
        ET-BERT  & 73.24 & 69.88 & 43.12 & 38.48 & 77.02 & 74.56 & 57.91 & 50.03 \\
        YaTC     & 72.31 & 68.65 & 21.79 & 16.33 & 78.55 & 74.90 & 44.48 & 27.71 \\
        NetMamba & 72.47 & 68.28 & 10.76 &  7.41 & 78.59 & 74.54 & 40.16 & 20.75 \\
        \textbf{Ours}
        & \textbf{78.53} & \textbf{72.14}
        & \textbf{73.56} & \textbf{67.89}
        & \textbf{83.93} & \textbf{82.19}
        & \textbf{76.03} & \textbf{73.02} \\
        \bottomrule
    \end{tabular*}
\end{table*}

\begin{table*}[!t]
    \centering
    \caption{Performance comparison under the random-based setting.}
    \label{tab:random_results}
    \small
    \setlength{\tabcolsep}{5.5pt}
    \renewcommand{\arraystretch}{1.15}
    \begin{tabular*}{\textwidth}{@{\extracolsep{\fill}}l
            S[table-format=2.2] S[table-format=2.2]
            S[table-format=2.2] S[table-format=2.2]
            S[table-format=2.2] S[table-format=2.2]
            S[table-format=2.2] S[table-format=2.2]
            @{}}
        \toprule
        \multirow{3}{*}{Method}
        & \multicolumn{4}{c}{CSTNET-APP(random split)}
        & \multicolumn{4}{c}{CSTNET-SERVICE(random split)} \\
        \cmidrule(lr){2-5} \cmidrule(lr){6-9}
        & \multicolumn{2}{c}{Unfrozen}
        & \multicolumn{2}{c}{Frozen}
        & \multicolumn{2}{c}{Unfrozen}
        & \multicolumn{2}{c}{Frozen} \\
        \cmidrule(lr){2-3} \cmidrule(lr){4-5}
        \cmidrule(lr){6-7} \cmidrule(lr){8-9}
        & {ACC (\%)} & {Macro-F1 (\%)}
        & {ACC (\%)} & {Macro-F1 (\%)}
        & {ACC (\%)} & {Macro-F1 (\%)}
        & {ACC (\%)} & {Macro-F1 (\%)} \\
        \midrule
        ET-BERT  & 91.52 & 90.17 & 53.50 & 48.31 & 94.15 & 93.88 & 61.91 & 55.12 \\
        YaTC     & 88.41 & 86.81 & 32.49 & 24.93 & 90.57 & 89.49 & 44.21 & 27.66 \\
        NetMamba & 89.73 & 88.21 & 21.51 & 13.93 & 91.63 & 90.36 & 40.89 & 23.86 \\
        \textbf{Ours}
        & \textbf{92.25} & \textbf{90.67}
        & \textbf{88.12} & \textbf{86.85}
        & \textbf{94.63} & \textbf{94.37}
        & \textbf{91.35} & \textbf{90.47} \\
        \bottomrule
    \end{tabular*}
\end{table*}

\begin{table*}[!t]
    \centering
    \caption{Ablation study under the time-based setting.}
    \label{tab:ablation_time}
    \small
    \setlength{\tabcolsep}{5.5pt}
    \renewcommand{\arraystretch}{1.15}
    \begin{tabular*}{\textwidth}{@{\extracolsep{\fill}}lcccccccc@{}}
        \toprule
        \multirow{3}{*}{Method}
        & \multicolumn{4}{c}{CSTNET-APP(time-based split)}
        & \multicolumn{4}{c}{CSTNET-SERVICE(time-based split)} \\
        \cmidrule(lr){2-5} \cmidrule(lr){6-9}
        & \multicolumn{2}{c}{Unfrozen}
        & \multicolumn{2}{c}{Frozen}
        & \multicolumn{2}{c}{Unfrozen}
        & \multicolumn{2}{c}{Frozen} \\
        \cmidrule(lr){2-3} \cmidrule(lr){4-5}
        \cmidrule(lr){6-7} \cmidrule(lr){8-9}
        & ACC (\%) & Macro-F1 (\%)
        & ACC (\%) & Macro-F1 (\%)
        & ACC (\%) & Macro-F1 (\%)
        & ACC (\%) & Macro-F1 (\%) \\
        \midrule
        Ours
        & 78.53 & 72.14
        & \textbf{73.56} & \textbf{67.89}
        & 83.93 & \textbf{82.19}
        & \textbf{76.03} & \textbf{73.02} \\

        Ours w/o Time Adv
        & 77.14 & 70.68
        & 69.31 & 64.41
        & 82.48 & 79.51
        & 75.42 & 72.14 \\

        Ours w/o Contrastive
        & 76.44 & 71.24
        & 64.12 & 60.05
        & \textbf{84.00} & 81.97
        & 74.29 & 69.77 \\

        Ours w/o Both
        & \textbf{79.76} & \textbf{74.67}
        & 61.12 & 57.34
        & 83.70 & 81.23
        & 71.32 & 66.48 \\
        \bottomrule
    \end{tabular*}
\end{table*}

\subsection{Online Evaluation}
\label{sec:exp_online}

\textbf{Online setting.}
The online policy-gradient hyperparameters are summarized in
Table~\ref{tab:online_hparams}. The policy and value networks are
three-layer MLPs with two 256-neuron hidden layers and rectified linear unit (ReLU)
activations. The classifier-head update is sensitive to the learning
rate under weak QoE feedback. Small values yield little adaptation,
and large values destabilize training. Hence we set the classifier-head
learning rate to $3\times10^{-5}$ for SEROA and ET-BERT,
$1\times10^{-4}$ for YaTC, and $3\times10^{-4}$ for NetMamba.

\begin{table}[H]
    \centering
    \caption{Hyperparameters of the online-learning module.}
    \label{tab:online_hparams}
    \small
    \setlength{\tabcolsep}{5pt}
    \renewcommand{\arraystretch}{1.08}
    \begin{tabular}{lc}
        \toprule
        \textbf{Hyperparameter}           & \textbf{Value}   \\
        \midrule
        Online adaptation episodes        & 150              \\
        Policy/value hidden widths        & 256, 256         \\
        Activation function               & ReLU             \\
        Multi-step updates $K$            & 8                \\
        PPO clipping parameter $\epsilon$ & 0.2              \\
        Value-loss weight $\lambda_v$     & 0.5              \\
        Entropy weight $\lambda_e$        & 0.01             \\
        Optimizer                         & Adam             \\
        Policy/value learning rate        & $3\times10^{-4}$ \\
        Classifier-head learning rate     & Method-specific  \\
        \bottomrule
    \end{tabular}
\end{table}

We conduct online evaluation on CSTNET-SERVICE under the
time-based split. The post-threshold portion is randomly divided
at a $1{:}1$ ratio into an online adaptation stream and a held-out
online test set. In addition to the Macro-F1 on these two portions,
we report the Macro-F1 on the original offline-training split to
measure forgetting during online adaptation.

We consider an urban macrocell single-cell downlink with $N=6$
active users, bandwidth $W=10$\,MHz, total transmit-power budget
$P_{\max}=1$\,W, and noise power $\sigma^2=10^{-8}$. The users
are distributed within a $0.05$--$0.30$\,km range from the BS. The large-scale fading for user $i$ $L_i$is generated from
the urban macrocell path-loss model in 3GPP Technical Report (TR)~36.942
\cite{TR36942_PathLoss}:
\begin{equation}
    L_i = 128.1 + 37.6\log_{10}(d_i),\qquad
    \beta_i = 10^{-L_i/10},
    \label{eq:exp_path_loss}
\end{equation}
where $d_i$ is measured in kilometers. The service profiles and
reference rates of the five CSTNET-SERVICE categories are given in
Table~\ref{tab:cstnet_service_profiles}.

\textbf{Online comparison.}
Following Section~\ref{sec:pg}, we freeze the encoder and update
the policy network $\pi_\omega$, the value network $V_\varphi$,
and the classifier head $h_\phi$ under QoE feedback. ET-BERT,
YaTC, NetMamba, and our method use the same RAN online-learning
framework and the same reward normalization. The reward also compares
the online policies with the static allocation references defined in
Section~\ref{sec:prelim_qoe}, including Oracle, Class-Agnostic, and the
predicted-class allocations driven by ET-BERT and our method.
For comparison, we normalize all reward curves by Oracle.

\begin{figure*}[!t]
    \centering
    \subfloat[Reward]{\includegraphics[width=0.245\textwidth]{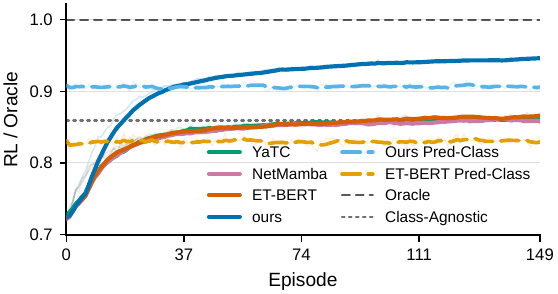}}\hfill
    \subfloat[Train Macro-F1]{\includegraphics[width=0.245\textwidth]{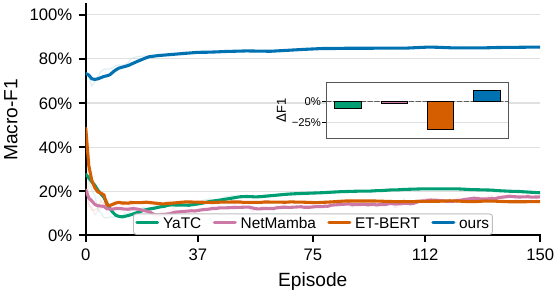}}\hfill
    \subfloat[Test Macro-F1]{\includegraphics[width=0.245\textwidth]{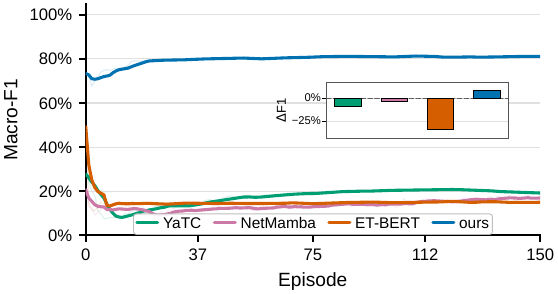}}\hfill
    \subfloat[Offline Train Macro-F1]{\includegraphics[width=0.245\textwidth]{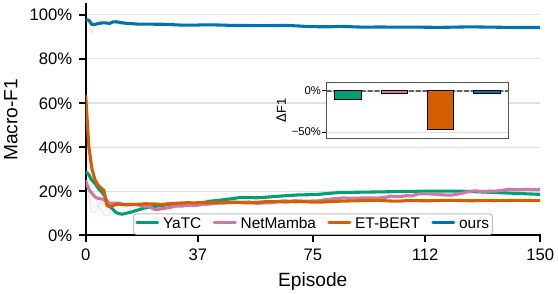}}
    \caption{Online-learning performance initialized from frozen fine-tuning. The reward panel includes online policies and static allocation references; the other panels report online-adaptation Macro-F1, held-out online-test Macro-F1, and offline-training Macro-F1 of the online learners.}
    \label{fig:online_perf_frozen}
\end{figure*}

\textbf{Initialization from frozen fine-tuning.}
Figure~\ref{fig:online_perf_frozen} reports the results for models
initialized by frozen fine-tuning. In the reward panel, Ours
Pred-Class is already well above ET-BERT Pred-Class and the
Class-Agnostic reference, showing that the frozen representation
provides reliable service information for static RAN allocation. The
online policy initialized from our method further improves the
oracle-normalized reward and moves closer to Oracle. The online
rewards of ET-BERT, YaTC, and NetMamba remain close to the
Class-Agnostic reference. Our method also maintains the highest
held-out online-test Macro-F1 throughout adaptation, whereas the
online-test and offline-training Macro-F1 scores of the comparison
methods decrease. These curves show that limited representation
performance makes the classifier head more susceptible to noisy
QoE feedback.

\begin{figure*}[!t]
    \centering
    \subfloat[Reward]{\includegraphics[width=0.245\textwidth]{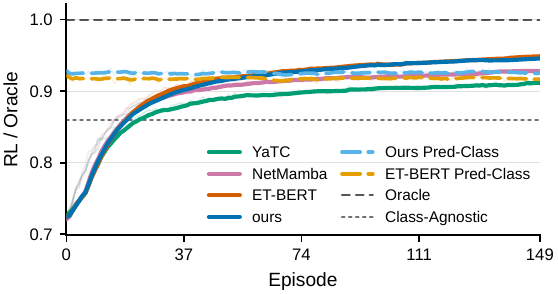}}\hfill
    \subfloat[Train Macro-F1]{\includegraphics[width=0.245\textwidth]{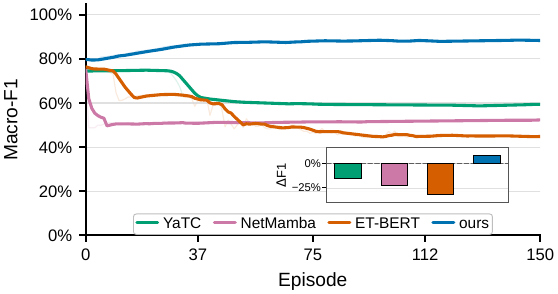}}\hfill
    \subfloat[Test Macro-F1]{\includegraphics[width=0.245\textwidth]{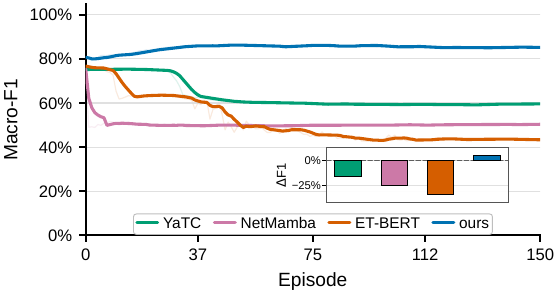}}\hfill
    \subfloat[Offline Train Macro-F1]{\includegraphics[width=0.245\textwidth]{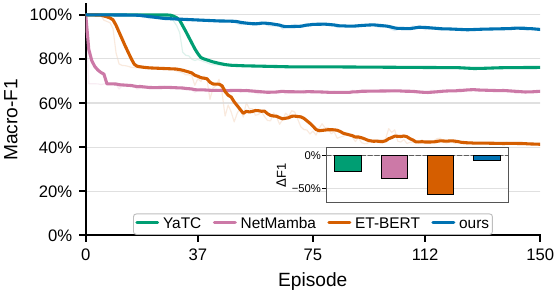}}
    \caption{Online-learning performance initialized from unfrozen fine-tuning. The reward panel includes online policies and static allocation references; the other panels report online-adaptation Macro-F1, held-out online-test Macro-F1, and offline-training Macro-F1 of the online learners.}
    \label{fig:online_perf_unfrozen}
\end{figure*}

\textbf{Initialization from unfrozen fine-tuning.}
Figure~\ref{fig:online_perf_unfrozen} reports the results for
models initialized by unfrozen fine-tuning. The initial offline
classification gap is narrower than in the frozen setting, and both
predicted-class references provide competitive static allocation
rewards above the Class-Agnostic reference. Our online policy still
attains the highest reward and remains closest to Oracle, indicating
that QoE-feedback adaptation improves power allocation beyond the
fixed predicted-class mapping. Our method also improves the held-out
online-test Macro-F1 while preserving its offline-training Macro-F1.
The classification curves of the comparison methods degrade during
adaptation. These results show that stable representations remain
important when online learning relies on noisy QoE feedback.

\begin{figure*}[!t]
    \centering
    \subfloat[Frozen Reward]{\includegraphics[width=0.245\textwidth]{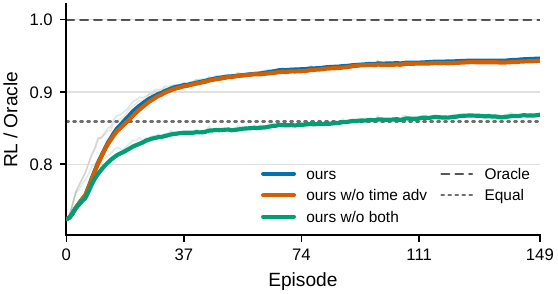}}\hfill
    \subfloat[Frozen Test Macro-F1]{\includegraphics[width=0.245\textwidth]{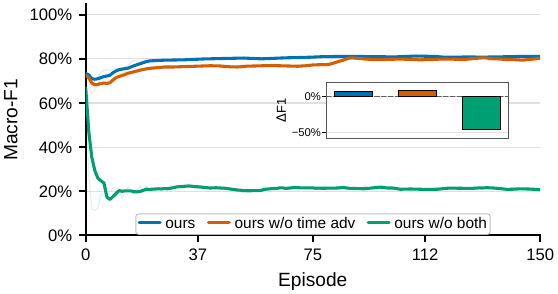}}\hfill
    \subfloat[Unfrozen Reward]{\includegraphics[width=0.245\textwidth]{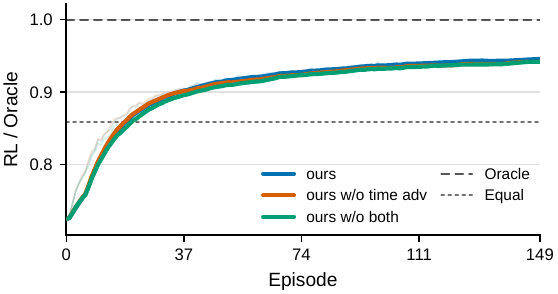}}\hfill
    \subfloat[Unfrozen Test Macro-F1]{\includegraphics[width=0.245\textwidth]{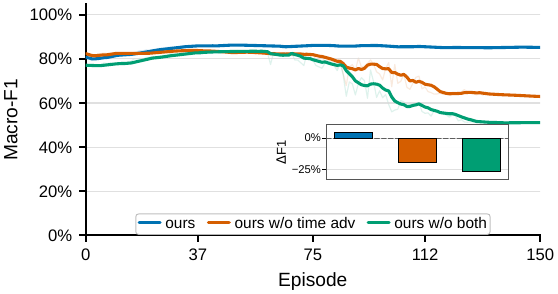}}
    \caption{Offline pretraining component ablation under frozen and unfrozen fine-tuning. The panels show oracle-normalized reward and held-out online-test Macro-F1.}
    \label{fig:offline_pretrain_ablation}
\end{figure*}

\textbf{Offline pretraining ablation.}
Figure~\ref{fig:offline_pretrain_ablation} evaluates the offline
pretraining components under both initialization settings. Adding
the contrastive objective to StableMLM substantially
improves the reward and held-out online-test Macro-F1. Adding the
time-adversarial objective provides a further gain. The same
ordering holds for both frozen and unfrozen fine-tuning. These
results confirm that robust offline representations improve
adaptation under noisy QoE feedback.

\begin{figure*}[!t]
    \centering
    \subfloat[Frozen Reward]{\includegraphics[width=0.245\textwidth]{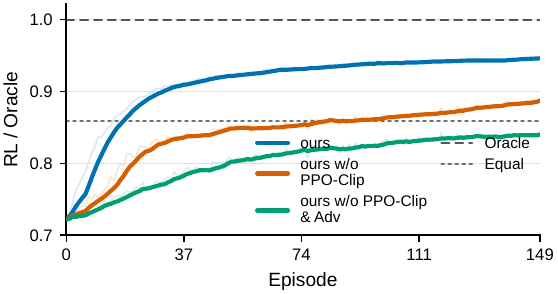}}\hfill
    \subfloat[Frozen Test Macro-F1]{\includegraphics[width=0.245\textwidth]{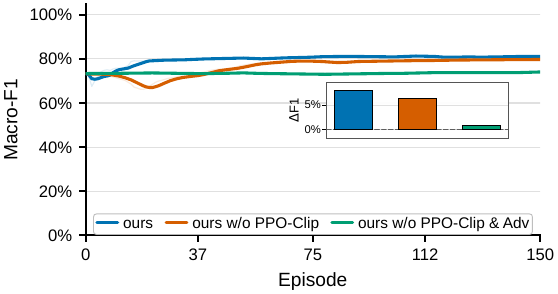}}\hfill
    \subfloat[Unfrozen Reward]{\includegraphics[width=0.245\textwidth]{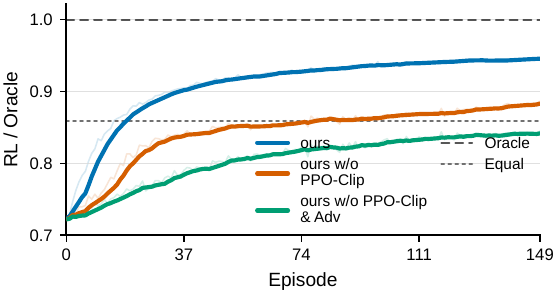}}\hfill
    \subfloat[Unfrozen Test Macro-F1]{\includegraphics[width=0.245\textwidth]{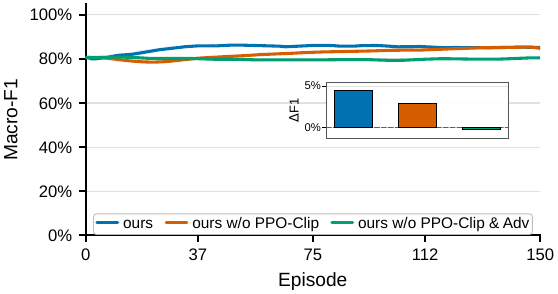}}
    \caption{Online policy-gradient component ablation under frozen and unfrozen fine-tuning. The panels show oracle-normalized reward and held-out online-test Macro-F1.}
    \label{fig:online_pg_ablation}
\end{figure*}

\textbf{Online policy-gradient ablation.}
Figure~\ref{fig:online_pg_ablation} evaluates the advantage
baseline and PPO clipped surrogate. The full method converges
faster and attains the highest reward and held-out online-test
Macro-F1 under both initialization settings. Removing PPO clipping reduces performance, while removing the advantage baseline causes a further decline.

\subsection{Inference Efficiency}
\label{sec:exp_efficiency}

To evaluate the computational overhead of the proposed method, we
measure pretraining and inference efficiency. Table~\ref{tab:inference_efficiency} reports the number of
trainable and total parameters, peak memory, per-sample latency,
and throughput with batch size $32$ on a
single GPU. Compared with the baseline pretraining stage, our
pretraining requires higher memory and latency because it jointly
optimizes StableMLM, contrastive learning and the time-adversarial branch.
However, this
additional cost is incurred only in the offline stage before
deployment and will not influence online inference.

At deployment time, the classifier is invoked once for
each early-flow sample, and the predicted service class is retained
throughout the flow lifetime for QoS mapping and scheduling. Therefore,
the deployed classifier does not require per-slot inference. The
inference latency is $0.64$\,ms per sample, corresponding to a
throughput of $1563.4$ samples/s, which is sufficient for real-time
flow-level service identification in the considered service-aware
mobile-network setting.

\begin{table}[t]
\centering
\caption{Pretraining overhead and inference efficiency.}
\label{tab:inference_efficiency}
\small
\setlength{\tabcolsep}{2.0pt}
\renewcommand{\arraystretch}{1.08}
\begin{tabular}{@{}l
c
S[table-format=1.2]
S[table-format=1.2]
S[table-format=4.1]
@{}}
\toprule
\textbf{Stage}    & \textbf{Params}        & {\textbf{Mem.}} & {\textbf{Time}}        & {\textbf{Throughput}}  \\
& {\textbf{Train/Total}} & {\textbf{(GB)}} & {\textbf{(ms/sample)}} & {\textbf{(samples/s)}} \\
\midrule
Baseline pretrain & 86.2M/86.2M            & 5.29            & 2.18                   & 458.7                  \\
Ours pretrain     & 86.9M/173.9M           & 9.81            & 4.88                   & 205.1                  \\
Inference         & 86.8M/86.8M            & 1.00            & 0.64                   & 1563.4                 \\
\bottomrule
\end{tabular}
\end{table}

%% file: refs.bib
@IEEEtranBSTCTL{IEEEtranBSTCTL_NoDash,
  CTLdash_repeated_names = "no"
}

@InProceedings{moore2005toward,
author="Moore, Andrew W.
and Papagiannaki, Konstantina",
editor="Dovrolis, Constantinos",
title="Toward the Accurate Identification of Network Applications",
booktitle="Passive and Active Network Measurement",
year="2005",
publisher="Springer Berlin Heidelberg",
address="Berlin, Heidelberg",
pages="41--54",
isbn="978-3-540-31966-5"
}

@article{Crotti2007traffic,
author = {Crotti, Manuel and Dusi, Maurizio and Gringoli, Francesco and Salgarelli, Luca},
title = {Traffic classification through simple statistical fingerprinting},
year = {2007},
issue_date = {January 2007},
publisher = {Association for Computing Machinery},
address = {New York, NY, USA},
volume = {37},
number = {1},
issn = {0146-4833},
url = {https://doi.org/10.1145/1198255.1198257},
doi = {10.1145/1198255.1198257},
journal = {SIGCOMM Comput. Commun. Rev.},
month = jan,
pages = {5–-16},
numpages = {12}
}

@inproceedings{Bernaille2006early,
author = {Bernaille, Laurent and Teixeira, Renata and Salamatian, Kave},
title = {Early application identification},
year = {2006},
isbn = {1595934561},
publisher = {Association for Computing Machinery},
address = {New York, NY, USA},
url = {https://doi.org/10.1145/1368436.1368445},
doi = {10.1145/1368436.1368445},
booktitle = {Proceedings of the 2006 ACM CoNEXT Conference},
articleno = {6},
numpages = {12},
location = {Lisboa, Portugal},
series = {CoNEXT '06}
}

@ARTICLE{Taylor2018Robust,
  author={Taylor, Vincent F. and Spolaor, Riccardo and Conti, Mauro and Martinovic, Ivan},
  journal={IEEE Transactions on Information Forensics and Security}, 
  title={Robust Smartphone App Identification via Encrypted Network Traffic Analysis}, 
  year={2018},
  volume={13},
  number={1},
  pages={63-78},
  doi={10.1109/TIFS.2017.2737970}}

@INPROCEEDINGS{Taylor2016AppScanner,
  author={Taylor, Vincent F. and Spolaor, Riccardo and Conti, Mauro and Martinovic, Ivan},
  booktitle={2016 IEEE European Symposium on Security and Privacy (EuroS\&P)}, 
  title={{AppScanner}: Automatic Fingerprinting of Smartphone Apps from Encrypted Network Traffic}, 
  year={2016},
  volume={},
  number={},
  pages={439-454},
  doi={10.1109/EuroSP.2016.40}}

@article{Lotfollahi2020DeepPacket,
  author  = {Lotfollahi, Mohammad and Jafari Siavoshani, Mahdi and Shirali Hossein Zade, Ramin and Saberian, Mohammdsadegh},
  title   = {{Deep Packet}: a novel approach for encrypted traffic classification using deep learning},
  journal = {Soft Computing},
  volume  = {24},
  pages   = {1999--2012},
  year    = {2020},
  doi     = {10.1007/s00500-019-04030-2}
}

@ARTICLE{Rezaei2020MobileAppDL,
  author={Rezaei, Shahbaz and Kroencke, Bryce and Liu, Xin},
  journal={IEEE Access}, 
  title={Large-Scale Mobile App Identification Using Deep Learning}, 
  year={2020},
  volume={8},
  number={},
  pages={348-362},
  doi={10.1109/ACCESS.2019.2962018}}

@ARTICLE{Shapira2021FlowPic,
  author={Shapira, Tal and Shavitt, Yuval},
  journal={IEEE Transactions on Network and Service Management}, 
  title={{FlowPic}: A Generic Representation for Encrypted Traffic Classification and Applications Identification}, 
  year={2021},
  volume={18},
  number={2},
  pages={1218-1232},
  doi={10.1109/TNSM.2021.3071441}}

@inproceedings{Horowicz2022few,
author = {Horowicz, Eyal and Shapira, Tal and Shavitt, Yuval},
title = {A few shots traffic classification with {mini-FlowPic} augmentations},
year = {2022},
isbn = {9781450392594},
publisher = {Association for Computing Machinery},
address = {New York, NY, USA},
url = {https://doi.org/10.1145/3517745.3561436},
doi = {10.1145/3517745.3561436},
booktitle = {Proceedings of the 22nd ACM Internet Measurement Conference},
pages = {647–-654},
numpages = {8},
location = {Nice, France},
series = {IMC '22}
}

@inproceedings{Lin2022ETBERT,
author = {Lin, Xinjie and Xiong, Gang and Gou, Gaopeng and Li, Zhen and Shi, Junzheng and Yu, Jing},
title = {{ET-BERT}: A Contextualized Datagram Representation with Pre-training Transformers for Encrypted Traffic Classification},
year = {2022},
isbn = {9781450390965},
publisher = {Association for Computing Machinery},
address = {New York, NY, USA},
url = {https://doi.org/10.1145/3485447.3512217},
doi = {10.1145/3485447.3512217},
booktitle = {Proceedings of the ACM Web Conference 2022},
pages = {633–-642},
numpages = {10},
location = {Virtual Event, Lyon, France},
series = {WWW '22}
}

@article{Zhao2023YaTC,
  title={Yet Another Traffic Classifier: A Masked Autoencoder Based Traffic Transformer with Multi-Level Flow Representation},
  author={Zhao, Ruijie and Zhan, Mingwei and Deng, Xianwen and Wang, Yanhao and Wang, Yijun and Gui, Guan and Xue, Zhi},
  journal={Proceedings of the AAAI Conference on Artificial Intelligence},
  volume={37},
  number={4},
  pages={5420--5427},
  year={2023},
  publisher={AAAI Press},
  url={https://ojs.aaai.org/index.php/AAAI/article/view/25674},
  doi={10.1609/aaai.v37i4.25674}
}

@INPROCEEDINGS{Wang2024NetMamba,
  author={Wang, Tongze and Xie, Xiaohui and Wang, Wenduo and Wang, Chuyi and Zhao, Youjian and Cui, Yong},
  booktitle={2024 IEEE 32nd International Conference on Network Protocols (ICNP)}, 
  title={{NetMamba}: Efficient Network Traffic Classification Via Pre-Training Unidirectional {Mamba}}, 
  year={2024},
  volume={},
  number={},
  pages={1-11},
  doi={10.1109/ICNP61940.2024.10858569}}

@article{Chen2025MIETT,
  title={{MIETT}: Multi-Instance Encrypted Traffic Transformer for Encrypted Traffic Classification},
  volume={39},
  url={https://ojs.aaai.org/index.php/AAAI/article/view/33748},
  DOI={10.1609/aaai.v39i15.33748},
  abstractNote={Network traffic includes data transmitted across a network, such as web browsing and file transfers, and is organized into packets (small units of data) and flows (sequences of packets exchanged between two endpoints). Classifying encrypted traffic is essential for detecting security threats and optimizing network management. Recent advancements have highlighted the superiority of foundation models in this task, particularly for their ability to leverage large amounts of unlabeled data and demonstrate strong generalization to unseen data. However, existing methods that focus on token-level relationships fail to capture broader flow patterns, as tokens, defined as sequences of hexadecimal digits, typically carry limited semantic information in encrypted traffic. These flow patterns, which are crucial for traffic classification, arise from the interactions between packets within a flow, not just their internal structure. To address this limitation, we propose a Multi-Instance Encrypted Traffic Transformer (MIETT), which adopts a multi-instance approach where each packet is treated as a distinct instance within a larger bag representing the entire flow. This enables the model to capture both token-level and packet-level relationships more effectively through Two-Level Attention (TLA) layers, improving the model’s ability to learn complex packet dynamics and flow patterns. We further enhance the model’s understanding of temporal and flow-specific dynamics by introducing two novel pre-training tasks: Packet Relative Position Prediction (PRPP) and Flow Contrastive Learning (FCL). After fine-tuning, MIETT achieves state-of-the-art (SOTA) performance across five datasets, demonstrating its effectiveness in classifying encrypted traffic and understanding complex network behaviors.},
  number={15},
  journal={Proceedings of the AAAI Conference on Artificial Intelligence},
  author={Chen, Xu-Yang and Han, Lu and Zhan, De-Chuan and Ye, Han-Jia},
  year={2025},
  month={Apr.},
  pages={15922–-15929}
}

@misc{liu2025FlowletFormer,
      title={{FlowletFormer}: Network Behavioral Semantic Aware Pre-training Model for Traffic Classification}, 
      author={Liming Liu and Ruoyu Li and Qing Li and Meijia Hou and Yong Jiang and Mingwei Xu},
      year={2025},
      eprint={2508.19924},
      archivePrefix={arXiv},
      primaryClass={cs.LG},
      url={https://arxiv.org/abs/2508.19924}, 
}

@article{Navid2023DataDrift,
title = {Deep learning for encrypted traffic classification in the face of data drift: An empirical study},
journal = {Computer Networks},
volume = {225},
pages = {109648},
year = {2023},
issn = {1389-1286},
doi={10.1016/j.comnet.2023.109648},
url = {https://www.sciencedirect.com/science/article/pii/S1389128623000932},
author = {Navid Malekghaini and Elham Akbari and Mohammad A. Salahuddin and Noura Limam and Raouf Boutaba and Bertrand Mathieu and Stephanie Moteau and Stephane Tuffin}
}

@inproceedings{Zhao2025SweetDanger,
author = {Zhao, Yuqi and Dettori, Giovanni and Boffa, Matteo and Vassio, Luca and Mellia, Marco},
title = {The Sweet Danger of Sugar: Debunking Representation Learning for Encrypted Traffic Classification},
year = {2025},
isbn = {9798400715242},
publisher = {Association for Computing Machinery},
address = {New York, NY, USA},
url = {https://doi.org/10.1145/3718958.3750498},
doi = {10.1145/3718958.3750498},
booktitle = {Proceedings of the ACM SIGCOMM 2025 Conference},
pages = {296–-310},
numpages = {15},
location = {S\~{a}o Francisco Convent, Coimbra, Portugal},
series = {SIGCOMM '25}
}

@ARTICLE{Malekghaini2024AutoML4ETC,
  author={Malekghaini, Navid and Akbari, Elham and Salahuddin, Mohammad A. and Limam, Noura and Boutaba, Raouf and Mathieu, Bertrand and Moteau, Stephanie and Tuffin, Stephane},
  journal={IEEE Transactions on Network and Service Management}, 
  title={{AutoML4ETC}: Automated Neural Architecture Search for Real-World Encrypted Traffic Classification}, 
  year={2024},
  volume={21},
  number={3},
  pages={2715-2730},
  doi={10.1109/TNSM.2023.3324936}
}

@INPROCEEDINGS{Wickramasinghe2025SoK,
  author={Wickramasinghe, Nimesha and Shaghaghi, Arash and Tsudik, Gene and Jha, Sanjay},
  booktitle={2025 IEEE Symposium on Security and Privacy (SP)}, 
  title={{SoK}: Decoding the Enigma of Encrypted Network Traffic Classifiers},
  year={2025},
  volume={},
  number={},
  pages={1825-1843},
  doi={10.1109/SP61157.2025.00165}}

@article{Mohammady2021TraceAnonymization,
  author  = {Mohammady, Mehdi and Oqaily, Mohammad and Wadhwa, Bhaskar and Hong, Yuan and Lin, Dan and Poor, H. Vincent and Dowsley, Rafael},
  title   = {A Multi-view Approach to Preserve Privacy and Utility in Network Trace Anonymization},
  journal = {ACM Transactions on Privacy and Security},
  volume  = {24},
  number  = {3},
  articleno = {14},
  pages   = {1--36},
  year    = {2021},
  doi     = {10.1145/3439732}
}

@article{Cui2025FGSAT,
  author  = {Cui, Susu and Han, Xueying and Han, Dongqi and Wang, Zhiliang and Wang, Weihang and Jiang, Bo and Liu, Baoxu and Lu, Zhigang},
  title   = {{FG-SAT}: Efficient Flow Graph for Encrypted Traffic Classification Under Environment Shifts},
  journal = {IEEE Transactions on Information Forensics and Security},
  volume  = {20},
  pages   = {5326--5339},
  year    = {2025},
  doi     = {10.1109/TIFS.2025.3571663}
}

@misc{chen2025SelfEvolvingETC,
      title={Drift-oriented Self-evolving Encrypted Traffic Application Classification for Actual Network Environment}, 
      author={Zihan Chen and Guang Cheng and Jinhui Li and Tian Qin and Yuyang Zhou and Xing Luan},
      year={2025},
      eprint={2501.04246},
      archivePrefix={arXiv},
      primaryClass={cs.NI},
      url={https://arxiv.org/abs/2501.04246}, 
}

@ARTICLE{Lin2025ETooL,
  author={Lin, Xinjie and Xiong, Gang and Gou, Gaopeng and Dong, Wenqi and Yu, Jing and Li, Zhen and Xia, Wei},
  journal={IEEE Transactions on Information Forensics and Security}, 
  title={Respond to Change With Constancy: Instruction-Tuning With {LLM} for {Non-I.I.D.} Network Traffic Classification}, 
  year={2025},
  volume={20},
  number={},
  pages={5758-5773},
  doi={10.1109/TIFS.2025.3574971}}

@techreport{TS22261_ServiceReq,
  author      = {{3GPP}},
  title       = {{Service requirements for the 5G system; Stage 1}},
  institution = {{3rd Generation Partnership Project (3GPP)}},
  type        = {{Technical Specification TS 22.261, version 20.5.0, Release 20}},
  month       = dec,
  year        = {2025},
  url         = {https://www.3gpp.org/ftp/Specs/archive/22_series/22.261/22261-k50.zip}
}

@techreport{TS23501_QoS,
  author      = {{3GPP}},
  title       = {{System architecture for the 5G System (5GS); Stage 2}},
  institution = {{3rd Generation Partnership Project (3GPP)}},
  type        = {{Technical Specification TS 23.501, version 20.1.0, Release 20}},
  month       = mar,
  year        = {2026},
  url         = {https://www.3gpp.org/ftp/Specs/archive/23_series/23.501/23501-k10.zip}
}

@techreport{TS38300_NGRAN,
  author      = {{3GPP}},
  title       = {{NR; NR and NG-RAN Overall Description; Stage 2}},
  institution = {{3rd Generation Partnership Project (3GPP)}},
  type        = {{Technical Specification TS 38.300, version 19.2.0, Release 19}},
  month       = mar,
  year        = {2026},
  url         = {https://www.3gpp.org/ftp/Specs/archive/38_series/38.300/38300-j20.zip}
}

@techreport{TS37340_MRDC,
  author      = {{3GPP}},
  title       = {{Evolved Universal Terrestrial Radio Access (E-UTRA) and NR; Multi-connectivity; Overall Description; Stage 2}},
  institution = {{3rd Generation Partnership Project (3GPP)}},
  type        = {{Technical Specification TS 37.340, version 19.0.0, Release 19}},
  month       = sep,
  year        = {2025},
  url         = {https://www.3gpp.org/ftp/Specs/archive/37_series/37.340/37340-j00.zip}
}

@techreport{TS23503_PCC,
  author      = {{3GPP}},
  title       = {{Policy and charging control framework for the 5G System (5GS); Stage 2}},
  institution = {{3rd Generation Partnership Project (3GPP)}},
  type        = {{Technical Specification TS 23.503, version 20.1.0, Release 20}},
  month       = mar,
  year        = {2026},
  url         = {https://www.3gpp.org/ftp/Specs/archive/23_series/23.503/23503-k10.zip}
}

@techreport{TS28405_QMC,
  author      = {{3GPP}},
  title       = {{Quality of Experience (QoE) measurement collection; Control and configuration}},
  institution = {{3rd Generation Partnership Project (3GPP)}},
  type        = {{Technical Specification TS 28.405, version 19.0.0, Release 19}},
  month       = jun,
  year        = {2024},
  url         = {https://www.3gpp.org/ftp/Specs/archive/28_series/28.405/28405-j00.zip}
}

@techreport{TS28105_AIML,
  author      = {{3GPP}},
  title       = {{Management and orchestration; Artificial Intelligence / Machine Learning (AI/ML) management}},
  institution = {{3rd Generation Partnership Project (3GPP)}},
  type        = {{Technical Specification TS 28.105, version 19.5.0, Release 19}},
  month       = mar,
  year        = {2026},
  url         = {https://www.3gpp.org/ftp/Specs/archive/28_series/28.105/28105-j50.zip}
}

@techreport{ETSI_MEC012_RNIS,
  author      = {{ETSI}},
  title       = {{Multi-access Edge Computing (MEC); Radio Network Information API}},
  institution = {{European Telecommunications Standards Institute (ETSI)}},
  type        = {{Group Specification}},
  number      = {{ETSI GS MEC 012}},
  year        = {2019},
  note        = {Version 2.1.1},
  url         = {https://www.etsi.org/deliver/etsi_gs/MEC/001_099/012/02.01.01_60/gs_MEC012v020101p.pdf}
}

@techreport{TR26928_XR5G,
  author      = {{3GPP}},
  title       = {{Extended Reality (XR) in 5G}},
  institution = {{3rd Generation Partnership Project (3GPP)}},
  type        = {{Technical Report TR 26.928, version 19.0.0, Release 19}},
  month       = oct,
  year        = {2025},
  url         = {https://www.3gpp.org/ftp/Specs/archive/26_series/26.928/26928-j00.zip}
}

@INPROCEEDINGS{Fei2023RealTime,
  author={Fei, Xiao and Martins, Philippe and Lu, Jialiang},
  booktitle={2023 IEEE 98th Vehicular Technology Conference (VTC2023-Fall)}, 
  title={Real-time Traffic Classification for {5G} {NSA} Encrypted Data Flows With Physical Channel Records}, 
  year={2023},
  volume={},
  number={},
  pages={1-6},
  doi={10.1109/VTC2023-Fall60731.2023.10333518}}

@article{Rezaei2019DeepLearningETC,
  author  = {Shahbaz Rezaei and Xin Liu},
  title   = {Deep Learning for Encrypted Traffic Classification: An Overview},
  journal = {IEEE Communications Magazine},
  volume  = {57},
  number  = {5},
  pages   = {76--81},
  year    = {2019},
  doi     = {10.1109/MCOM.2019.1800819}
}

@misc{rfc8446,
  series = {Request for Comments},
  number = 8446,
  howpublished = {RFC 8446},
  publisher = {RFC Editor},
  doi = {10.17487/RFC8446},
  url = {https://www.rfc-editor.org/info/rfc8446},
  author = {Eric Rescorla},
  title = {{The Transport Layer Security (TLS) Protocol Version 1.3}},
  pagetotal = 160,
  year = 2018,
  month = aug
}

@inproceedings{Devlin2019BERT,
  title = {{BERT}: Pre-training of Deep Bidirectional Transformers for Language Understanding},
  author = {Devlin, Jacob and Chang, Ming-Wei and Lee, Kenton and Toutanova, Kristina},
  booktitle = {Proceedings of the 2019 Conference of the North American Chapter of the Association for Computational Linguistics: Human Language Technologies},
  pages = {4171--4186},
  year = {2019},
  publisher = {Association for Computational Linguistics},
  doi = {10.18653/v1/N19-1423},
  url = {https://aclanthology.org/N19-1423/}
}

@inproceedings{He2022MAE,
  author = {He, Kaiming and Chen, Xinlei and Xie, Saining and Li, Yanghao and Doll{\'a}r, Piotr and Girshick, Ross},
  title = {Masked Autoencoders Are Scalable Vision Learners},
  booktitle = {Proceedings of the IEEE/CVF Conference on Computer Vision and Pattern Recognition (CVPR)},
  pages = {16000--16009},
  year = {2022}
}

@inproceedings{He2020MoCo,
  author = {He, Kaiming and Fan, Haoqi and Wu, Yuxin and Xie, Saining and Girshick, Ross},
  title = {Momentum Contrast for Unsupervised Visual Representation Learning},
  booktitle = {Proceedings of the IEEE/CVF Conference on Computer Vision and Pattern Recognition (CVPR)},
  pages = {9729--9738},
  year = {2020}
}

@misc{Chen2020MoCov2,
  title = {Improved Baselines with Momentum Contrastive Learning},
  author = {Chen, Xinlei and Fan, Haoqi and Girshick, Ross and He, Kaiming},
  year = {2020},
  eprint = {2003.04297},
  archivePrefix = {arXiv},
  primaryClass = {cs.CV},
  doi = {10.48550/arXiv.2003.04297},
  url = {https://arxiv.org/abs/2003.04297}
}

@misc{Oord2018CPC,
  title = {Representation Learning with Contrastive Predictive Coding},
  author = {Oord, Aaron van den and Li, Yazhe and Vinyals, Oriol},
  year = {2018},
  eprint = {1807.03748},
  archivePrefix = {arXiv},
  primaryClass = {cs.LG},
  doi = {10.48550/arXiv.1807.03748},
  url = {https://arxiv.org/abs/1807.03748}
}

@article{Ganin2016DANN,
  author  = {Yaroslav Ganin and Evgeniya Ustinova and Hana Ajakan and Pascal Germain and Hugo Larochelle and Fran{\c{c}}ois Laviolette and Mario Marchand and Victor Lempitsky},
  title   = {Domain-Adversarial Training of Neural Networks},
  journal = {Journal of Machine Learning Research},
  volume  = {17},
  number  = {59},
  pages   = {1--35},
  year    = {2016},
  url     = {https://jmlr.org/papers/v17/15-239.html}
}

@inproceedings{Li2010ContextualBandit,
  author    = {Lihong Li and Wei Chu and John Langford and Robert E. Schapire},
  title     = {A Contextual-Bandit Approach to Personalized News Article Recommendation},
  booktitle = {Proceedings of the 19th International Conference on World Wide Web},
  pages     = {661--670},
  year      = {2010},
  publisher = {ACM},
  doi       = {10.1145/1772690.1772758},
  url       = {https://doi.org/10.1145/1772690.1772758}
}

@article{Williams1992REINFORCE,
  title = {Simple statistical gradient-following algorithms for connectionist reinforcement learning},
  volume = {8},
  issn = {1573-0565},
  url = {https://doi.org/10.1007/BF00992696},
  doi = {10.1007/BF00992696},
  number = {3-4},
  journal = {Machine Learning},
  publisher = {Springer Science and Business Media LLC},
  author = {Williams, Ronald J.},
  year = {1992},
  month = {May},
  pages = {229--256}
}

@misc{Schulman2017PPO,
  title = {Proximal Policy Optimization Algorithms},
  author = {Schulman, John and Wolski, Filip and Dhariwal, Prafulla and Radford, Alec and Klimov, Oleg},
  year = {2017},
  eprint = {1707.06347},
  archivePrefix = {arXiv},
  primaryClass = {cs.LG},
  doi = {10.48550/arXiv.1707.06347},
  url = {https://arxiv.org/abs/1707.06347}
}

@article{Laghari2019CloudGamingQoE,
  author  = {Asif Ali Laghari and Hui He and Kamran Ali Memon and Rashid Ali Laghari and Imtiaz Ali Halepoto and Asiya Khan},
  title   = {Quality of Experience ({QoE}) in Cloud Gaming Models: A Review},
  journal = {Multiagent and Grid Systems},
  volume  = {15},
  number  = {3},
  pages   = {289--304},
  year    = {2019},
  doi     = {10.3233/MGS-190313}
}

@article{Sousa2020QoEOrientedScheduling,
  author  = {Sousa, Ivo and Queluz, Maria Paula and Rodrigues, Ant{\'o}nio},
  title   = {A Survey on {QoE}-Oriented Wireless Resources Scheduling},
  journal = {Journal of Network and Computer Applications},
  volume  = {158},
  pages   = {102594},
  year    = {2020},
  doi     = {10.1016/j.jnca.2020.102594}
}

@INPROCEEDINGS{Elgabli2018QoERA,
  author={Elgabli, Anis and Elghariani, Ali and Aggarwal, Vaneet and Bell, Mark},
  booktitle={2018 IEEE Global Communications Conference (GLOBECOM)}, 
  title={{QoE}-Aware Resource Allocation for Small Cells}, 
  year={2018},
  volume={},
  number={},
  pages={1-6},
  doi={10.1109/GLOCOM.2018.8647828}}

@article{Ojaghi2022SlicedRAN,
  author  = {Behnam Ojaghi and Ferran Adelantado and Angelos Antonopoulos and Christos Verikoukis},
  title   = {{SlicedRAN}: Service-Aware Network Slicing Framework for {5G} Radio Access Networks},
  journal = {IEEE Systems Journal},
  volume  = {16},
  number  = {2},
  pages   = {2556--2567},
  year    = {2022},
  doi     = {10.1109/JSYST.2021.3064398}
}

@article{Mei2021RANSlicingDRL,
  author  = {Mei, Jie and Wang, Xianbin and Zheng, Kan and Boudreau, Gary and Bin Sediq, Akram and Abou-Zeid, Hatem},
  title   = {Intelligent Radio Access Network Slicing for Service Provisioning in {6G}: A Hierarchical Deep Reinforcement Learning Approach},
  journal = {IEEE Transactions on Communications},
  volume  = {69},
  number  = {9},
  pages   = {6063--6078},
  year    = {2021},
  doi     = {10.1109/TCOMM.2021.3090423}
}

@article{Polese2023ColORAN,
  author  = {Polese, Michele and Bonati, Leonardo and D'Oro, Salvatore and Basagni, Stefano and Melodia, Tommaso},
  title   = {{ColO-RAN}: Developing Machine Learning-Based {xApps} for Open {RAN} Closed-Loop Control on Programmable Experimental Platforms},
  journal = {IEEE Transactions on Mobile Computing},
  volume  = {22},
  number  = {10},
  pages   = {5787--5800},
  year    = {2023},
  doi     = {10.1109/TMC.2022.3188013}
}

@article{Bronzino2019EncryptedVideoQoE,
  author  = {Bronzino, Francesco and Schmitt, Paul and Ayoubi, Sara and Martins, Guilherme and Teixeira, Renata and Feamster, Nick},
  title   = {Inferring Streaming Video Quality from Encrypted Traffic: Practical Models and Deployment Experience},
  journal = {Proceedings of the ACM on Measurement and Analysis of Computing Systems},
  volume  = {3},
  number  = {3},
  articleno = {56},
  pages   = {1--25},
  year    = {2019},
  doi     = {10.1145/3366704}
}

@inproceedings{Dimopoulos2016EncryptedVideoQoE,
  author    = {Dimopoulos, Giorgos and Leontiadis, Ilias and Barlet-Ros, Pere and Papagiannaki, Konstantina},
  title     = {Measuring Video {QoE} from Encrypted Traffic},
  booktitle = {Proceedings of the 2016 ACM Internet Measurement Conference},
  series    = {IMC},
  pages     = {513--526},
  year      = {2016},
  publisher = {ACM},
  doi       = {10.1145/2987443.2987459}
}

@techreport{TR36942_PathLoss,
  author      = {{3GPP}},
  title       = {{Evolved Universal Terrestrial Radio Access (E-UTRA); Radio Frequency (RF) system scenarios}},
  institution = {{3rd Generation Partnership Project (3GPP)}},
  type        = {{Technical Report TR 36.942, version 17.0.0, Release 17}},
  month       = mar,
  year        = {2022},
  url         = {https://www.3gpp.org/ftp/Specs/archive/36_series/36.942/36942-h00.zip}
}

@techreport{ITU_G1010,
  author      = {{ITU-T}},
  title       = {{End-user Multimedia QoS Categories}},
  institution = {{International Telecommunication Union}},
  type        = {{Recommendation}},
  number      = {{G.1010}},
  year        = {2001},
  month       = nov,
  url         = {https://www.itu.int/rec/T-REC-G.1010-200111-I/en}
}

@techreport{RFC6817_LEDBAT,
  author      = {Shalunov, Stanislav and Hazel, Greg and Iyengar, Janardhan and Kuehlewind, Mirja},
  title       = {{Low Extra Delay Background Transport (LEDBAT)}},
  institution = {{Internet Engineering Task Force}},
  type        = {{RFC}},
  number      = {6817},
  year        = {2012},
  month       = dec,
  doi         = {10.17487/RFC6817},
  url         = {https://www.rfc-editor.org/rfc/rfc6817}
}

@misc{AppleHLSAuthoring,
  author       = {{Apple Inc.}},
  title        = {{HTTP Live Streaming (HLS) Authoring Specification for Apple Devices}},
  howpublished = {\url{https://developer.apple.com/documentation/http-live-streaming/hls-authoring-specification-for-apple-devices}},
  note         = {Accessed: Jun. 1, 2026}
}

@INPROCEEDINGS{Reichl2010WeberFechnerQoE,
  author={Reichl, Peter and Egger, Sebastian and Schatz, Raimund and D'Alconzo, Alessandro},
  booktitle={2010 IEEE International Conference on Communications},
  title={The Logarithmic Nature of {QoE} and the Role of the {Weber-Fechner} Law in {QoE} Assessment},
  year={2010},
  pages={1--5},
  doi={10.1109/ICC.2010.5501894}
}

@ARTICLE{Lin1991JSD,
  author={Lin, Jianhua},
  journal={IEEE Transactions on Information Theory},
  title={Divergence Measures Based on the Shannon Entropy},
  year={1991},
  volume={37},
  number={1},
  pages={145--151},
  doi={10.1109/18.61115}
}

@book{goldsmith2005wireless,
  author    = {Goldsmith, Andrea},
  title     = {Wireless Communications},
  publisher = {Cambridge University Press},
  year      = {2005},
  doi       = {10.1017/CBO9780511841224},
  isbn      = {9780511841224}
}

@techreport{TR37817_RANAI,
  author      = {{3GPP}},
  title       = {{Evolved Universal Terrestrial Radio Access (E-UTRA) and NR; Study on enhancement for Data Collection for NR and EN-DC}},
  institution = {{3rd Generation Partnership Project (3GPP)}},
  type        = {{Technical Report TR 37.817, version 17.0.0, Release 17}},
  month       = apr,
  year        = {2022},
  url         = {https://www.3gpp.org/ftp/Specs/archive/37_series/37.817/37817-h00.zip}
}

@ARTICLE{Liu2026LAMeTA,
  author={Liu, Yinqiu and Liu, Guangyuan and Wang, Jiacheng and Zhang, Ruichen and Niyato, Dusit and Sun, Geng and Xiong, Zehui and Han, Zhu},
  journal={IEEE Journal on Selected Areas in Communications}, 
  title={{LAMeTA}: Intent-Aware Agentic Network Optimization via a Large {AI} Model-Empowered Two-Stage Approach}, 
  year={2026},
  volume={44},
  number={},
  pages={2462-2478},
  doi={10.1109/JSAC.2025.3642840}}
